\documentclass{SCIS2026}

\begin{document}
\ArticleType{REVIEW}
\Year{2025}
\Month{January}
\Vol{68}
\No{1}
\DOI{}
\ArtNo{}
\ReceiveDate{}
\ReviseDate{}
\AcceptDate{}
\OnlineDate{}
\AuthorMark{}
\AuthorCitation{}

\title{A New Paradigm of 6G Networks: Proactive Channel Cognition and Reconfiguration}{A New Paradigm of 6G Networks: Proactive Channel Cognition and Reconfiguration}

\author[1]{Wenyan MA}{}
\author[1]{Zixiang REN}{}
\author[1]{Weitong ZHAI}{}
\author[1]{Ge YAN}{}
\author[2]{Lipeng ZHU}{{zhulp@bit.edu.cn}}
\author[3]{Zhenyu XIAO}{}
\author[1]{Rui ZHANG}{}


\address[1]{Department of Electrical and Computer Engineering, National University of Singapore, Singapore 117583, Singapore}
\address[2]{State Key Laboratory of CNS/ATM and the School of Interdisciplinary Science, Beijing Institute of Technology, Beijing 100081, China}
\address[3]{School of Electronic and Information Engineering, Beihang University, Beijing 100191, China}

\abstract{The sixth-generation (6G) wireless networks are expected to enable the deep integration of communication, sensing, computing, control, and intelligence in highly dynamic environments. This evolution drives a fundamental transition from conventional passive channel adaptation to proactive channel cognition and reconfiguration, wherein wireless channels are no longer regarded as uncontrollable propagation media but as network resources that can be learned, predicted, and actively reconfigured. This paper presents a comprehensive overview of this emerging paradigm. We first review channel cognition through the channel knowledge map (CKM) as a systematic framework for learning and exploiting channel characteristics across space, time, and frequency domain. The definitions, construction methods, and applications in wireless networks of CKMs are comprehensively reviewed. Building upon channel cognition, we then review channel reconfiguration technologies from two complementary perspectives: transceiver-side reconfiguration enabled by movable antennas (MAs) and environment-side reconfiguration enabled by intelligent reflecting surfaces (IRSs). For both MA- and IRS-enabled wireless systems, we review their architectures, performance advantages, and key design challenges. Finally, we discuss several promising research directions to inspire further innovations in this burgeoning field.}


\maketitle

\section{Introduction} \label{Sec_Intro} 

\subsection{Evolution of Mobile Networks} 

Mobile communications have evolved from voice-centric first-generation (1G)/second-generation (2G) systems to broadband third-generation (3G)/fourth-generation (4G) networks and multi-service fifth-generation (5G) networks. For 5G, International Mobile Telecommunications (IMT)-2020 specifies a downlink peak data rate of 20 Gbit/s and a 1-ms user-plane latency for ultra-reliable low-latency communications (URLLC). Toward 2030 and beyond, IMT-2030 further encompasses immersive communication, hyper-reliable and low-latency communication, massive communication, ubiquitous connectivity, artificial intelligence (AI) and communication, and integrated sensing and communication (ISAC). Its scenario-dependent research targets include peak data rates of 50-200 Gbit/s, connection densities of $10^{6}$-$10^{8}$ devices/km$^{2}$, air-interface latency of 0.1-1 ms, and positioning accuracy of 1-10 cm, together with new sensing- and AI-related capabilities \cite{new-001,new-002,new-003}. These requirements are directly constrained by wireless propagation conditions, where blockage and fading effects limit data rate and reliability, rapid channel variations increase estimation overhead and consume the latency budget, and dense access intensifies interference \cite{001,002,003,004,005}. Hence, a revisit of conventional network operation and a rethink of new communication paradigm are needed to fundamentally address these challenges.



\begin{figure}[t] 
	\centering 
	\captionsetup[subfloat]{captionskip=5pt} 
	\subfloat[Traditional passive channel adaptation]{
		\includegraphics[width=8.26cm]{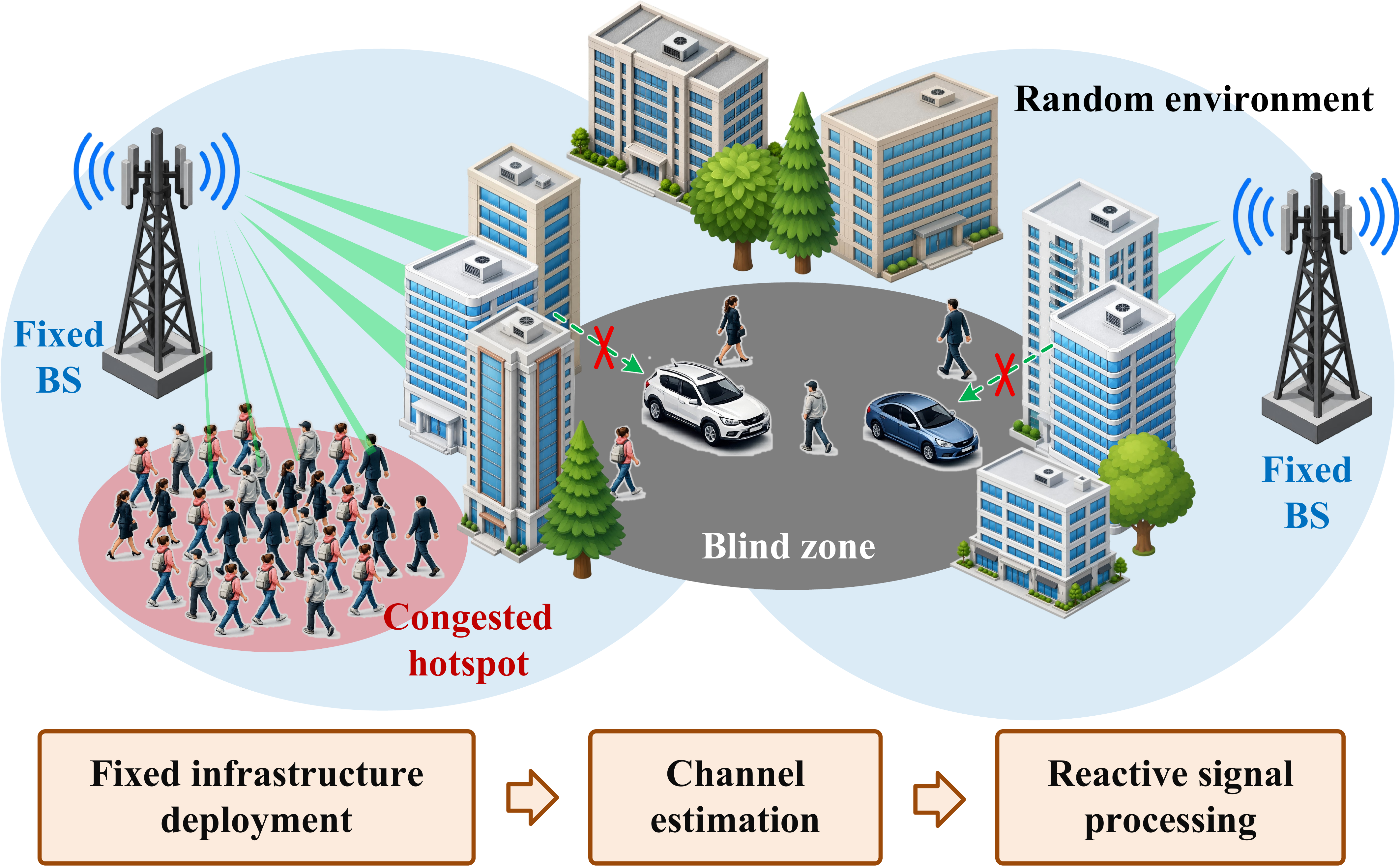} 
		\label{Fig1a} 
	}\hspace{0.2cm} 
	\subfloat[Proactive channel cognition and reconfiguration]{
		\includegraphics[width=8.26cm]{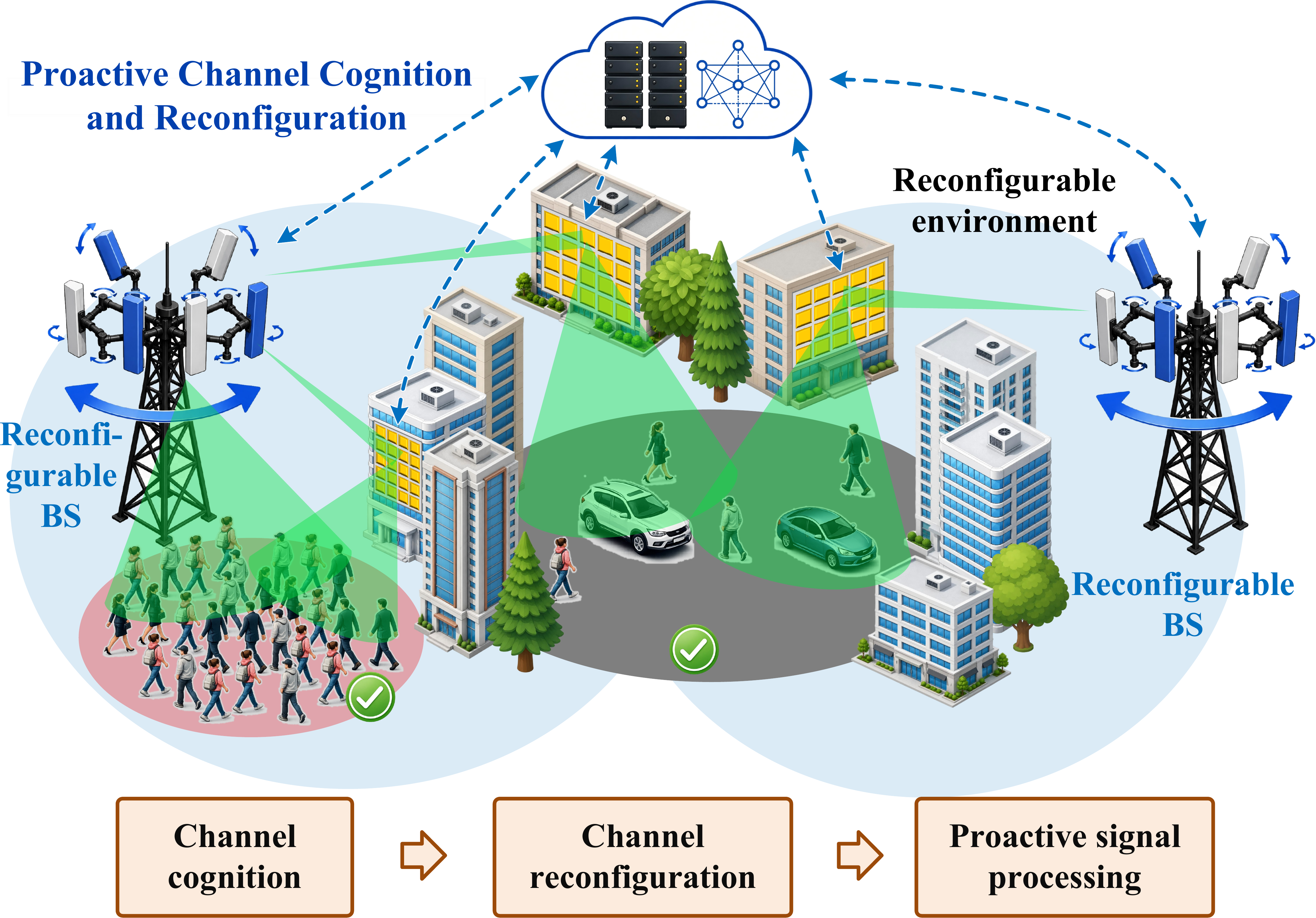} 
		\label{Fig1b} 
	}
	\caption{Illustration of the evolution from traditional passive channel adaptation to proactive channel cognition and reconfiguration. }
	\label{Fig1} 
	\vspace{-0.26cm} 
\end{figure}

Conventional wireless networks are typically scheduled and deployed based on long-term and coarse-grained channel statistics. Once deployed, base station (BS) locations and antenna configurations remain largely fixed, making the resulting signal propagation condition difficult to adjust over time. During network operation, instantaneous channel state information (CSI) is acquired through pilot-based channel estimation and feedback, after which the BS optimizes beamforming, power control, user scheduling, and resource allocation \cite{006,007,008,009}. This conventional ``estimate-and-adapt'' paradigm has shown to be effective in earlier cellular generations, but its limitations become increasingly pronounced in 6G scenarios. As illustrated in Fig. \ref{Fig1a}, fixed infrastructure cannot promptly accommodate random environments, mobile user clusters, or spatially nonuniform traffic demands, resulting in potential coverage blockages, overloaded regions, and underutilized resources. Moreover, the development trend of high-frequency transmission, extremely large-scale antenna arrays, ISAC, and space-air-ground-sea integrated networks substantially increases the overhead and complexity of channel acquisition, beam training, feedback, and real-time optimization. Consequently, purely reactive network operation based on instantaneous channel estimation is increasingly inefficient for meeting the low-overhead, low-latency, and high-throughput requirements in 6G. 


Fundamentally, the physical channel determines the information-theoretic performance ceiling of a wireless communication system. A cellular network can be abstractly modeled as a multiple-input multiple-output (MIMO) system, for which the channel capacity is given by
\begin{align} 
	\label{Shannon achievable rate} 
	C=B\log_{2}\det\left (  \mathbf{I}+\frac{1}{\sigma^{2}}\mathbf{H}\mathbf{Q}\mathbf{H}^{H} \right ) , 
\end{align} 
where $B$ is the system bandwidth, $\mathbf{H}$ denotes the channel matrix between all BSs' antennas and users' antennas, $\mathbf{Q}$ is the transmit covariance matrix, and $\sigma^{2}$ is the noise power. Conventional signal processing techniques, including channel coding, beamforming, and power allocation, seek to approach this performance limit by adapting transmission to a given channel $\mathbf{H}$ \cite{010,011,012,013}. However, when blockage, deep fading, or insufficient scattering weakens the channel conditions, such as channel gain and channel rank, the channel capacity becomes fundamentally limited. In such cases, passive adaptation can only exploit the available signal processing degrees of freedom (DoFs), but cannot directly create favorable channel conditions. Therefore, 6G network design should move beyond adapting transmission to a given channel toward understanding, predicting, and proactively improving the channel itself. Accordingly, the wireless channel should be regarded not merely as an external medium to be estimated and adapted to, but as a critical network resource that can be proactively cognized and reconfigured. 


\subsection{Paradigm Shift to Channel Cognition and Reconfiguration} 

The essence of the above paradigm shift is that wireless networks need to evolve from merely adapting to given channels toward channel cognition and channel reconfiguration. As illustrated in Fig.~\ref{Fig1b}, channel reconfiguration addresses how the channel can be improved through controllable mechanisms. Effective channel reconfiguration, however, requires the network to understand why a channel exhibits its current condition and how it may evolve, which is the role of channel cognition. In this sense, the wireless channel can gradually evolve from an external random propagation condition into a network resource that is understandable, predictable, and controllable. 


Channel reconfiguration seeks to improve wireless channels through controllable physical mechanisms. In a broad sense, it refers to techniques that modify the formation process of effective wireless channels by adjusting either transceiver structures or signal propagation environments. At the transceiver side, intelligent antenna technologies, such as movable antennas (MAs) \cite{010,shao2024Mag6DMA}, reconfigurable antennas \cite{026}, fluid antennas \cite{027}, pinching antennas \cite{liu2025pinching}, and rotatable antennas \cite{zheng2025rotatable}, can exploit additional spatial and electromagnetic DoFs by changing antenna positions, orientations, polarizations, radiation characteristics, or array structures. At the environment side, intelligent reflecting surfaces (IRSs)/reconfigurable intelligent surfaces (RISs) and their advanced variants, such as simultaneously transmitting and reflecting surfaces \cite{10133841}, beyond-diagonal IRS/RIS \cite{li2022beyond}, stacked intelligent metasurfaces \cite{an2024stacked}, movable intelligent surfaces \cite{zheng2025movableRIS}, and flexible intelligent metasurfaces \cite{an2025flexible}, can reshape propagation paths and energy distributions by controlling signal reflection, refraction, diffraction, scattering, and polarization states. In a broader sense, mobile BSs, unmanned-aerial-vehicle (UAV) platforms, deployable relays, and mobile access points (APs) can improve large-scale channel conditions by adapting infrastructure locations and service topologies, which can also be regarded as channel reconfiguration technologies \cite{033,034}. 

Effective channel reconfiguration requires the network to acquire and exploit sufficient knowledge of wireless channels, i.e., channel cognition. It covers a broad spectrum of techniques that acquire and exploit channel-related knowledge from different perspectives. For example, cognitive radio focuses on spectrum occupancy and interference awareness to support dynamic spectrum access \cite{014,015}. Radio maps provide spatial representations of received signal power, interference, spectrum utilization, and coverage states \cite{016,017}. Geometry-based and statistical channel models characterize propagation mechanisms such as path loss, shadowing, scattering, angular-delay dispersion, and spatio-temporal non-stationarity \cite{018,019}. Channel prediction leverages historical measurements, user mobility, and environmental dynamics to infer future channel states, while digital twin networks reproduce physical networks and propagation environments in virtual spaces for network evaluation and policy validation \cite{020,021}. More recently, AI and emerging generative or foundation models have further enhanced multi-source information fusion, cross-scenario reasoning, and generalized channel modeling and prediction \cite{022,023}. Establishing upon these advancements, the concept of channel knowledge map (CKM) has been introduced to characterize the mapping between spatial locations and channel features by extracting useful environmental knowledge, and thus can be regarded as a representative technique for implementing channel cognition \cite{zeng2024CKMtut,wu2024CKMbeamforming}. By transforming discrete channel measurements into reusable and predictable channel knowledge, it can support coverage evaluation, site deployment, beam management, enhanced channel acquisition, and proactive resource allocation.

Notably, channel cognition and channel reconfiguration should not be viewed as isolated research directions, but as two tightly coupled components of proactive channel management for 6G networks. Channel cognition provides knowledge and predictions of propagation environments, user distributions, and channel states, while channel reconfiguration exploits such information to adjust transceiver structures, propagation environments, or network deployment, thereby improving the effective channel condition. The measurements collected after channel reconfiguration can be further fed back to the cognition module to update channel knowledge, refine prediction models, and support subsequent decision-making. Through this cognition-reconfiguration-feedback loop, wireless networks are expected to evolve from conventional reactive systems into proactive intelligent systems capable of understanding, shaping, and continuously optimizing wireless channels. 

\subsection{Contributions and Organization} 


Existing studies have extensively investigated channel cognition \cite{021,zeng2024CKMtut,035,036,zeng2021CKM,038,039} and channel reconfiguration \cite{010,013,014,026,027,zheng2025rotatable,an2024stacked,ma2026survey}, laying important foundations for their respective areas. Nevertheless, most existing works are centered on specific technology categories or individual technical branches, while the closed-loop integration of channel cognition and channel reconfiguration as a unified paradigm has not been fully recognized in the literature. In particular, how channel knowledge can be translated into physical reconfiguration decisions, and how post-reconfiguration measurements can be fed back to update channel knowledge, have not yet been systematically reviewed under a unified proactive channel management framework. 

To fill this gap, this paper provides a systematic overview of proactive channel cognition and reconfiguration as a new paradigm for 6G wireless networks. The main contributions of this paper are summarized as follows. 

First, this paper establishes a unified cognition-reconfiguration framework for proactive channel management. Starting from 6G service requirements and wireless channel bottlenecks, we analyze the limitations of conventional passive channel adaptation and reveal the intrinsic connection between channel cognition and reconfiguration.


Second, this paper reviews channel cognition with a particular focus on CKMs. We summarize their fundamentals, construction methods, and representative applications, and briefly discuss related channel cognition technologies. 

Third, this paper reviews transceiver-side channel reconfiguration technologies, with a focus on MAs. We summarize how position, orientation, and structural DoFs of antennas can be exploited to reshape effective channels and discusses key issues in modeling, channel acquisition, and antenna movement optimization and scheduling.

Fourth, this paper reviews environment-side channel reconfiguration technologies represented by IRSs. We discuss how IRSs can reshape propagation environments for coverage enhancement and interference suppression, together with key challenges in deployment, channel acquisition, and joint optimization.

Finally, this paper identifies future research directions for proactive channel cognition and reconfiguration, including full-scenario channel cognition, collaborative channel reconfiguration, and embodied AI network architectures based on the cognition-reconfiguration-feedback loop. 

The remainder of this paper is organized as follows. Section II introduces channel cognition based on CKMs. Section III discusses transceiver-side channel reconfiguration based on MAs. Section IV presents environment-side channel reconfiguration based on IRSs. Section V discusses future research directions for proactive channel cognition and reconfiguration. Section VI concludes this paper.

\section{Channel Cognition through CKM}  \label{Sec_CKM}

\begin{figure*}[!t]
	\centering
	\captionsetup[subfloat]{captionskip=3pt}
	\subfloat[]{
		\includegraphics[width=0.98\textwidth]{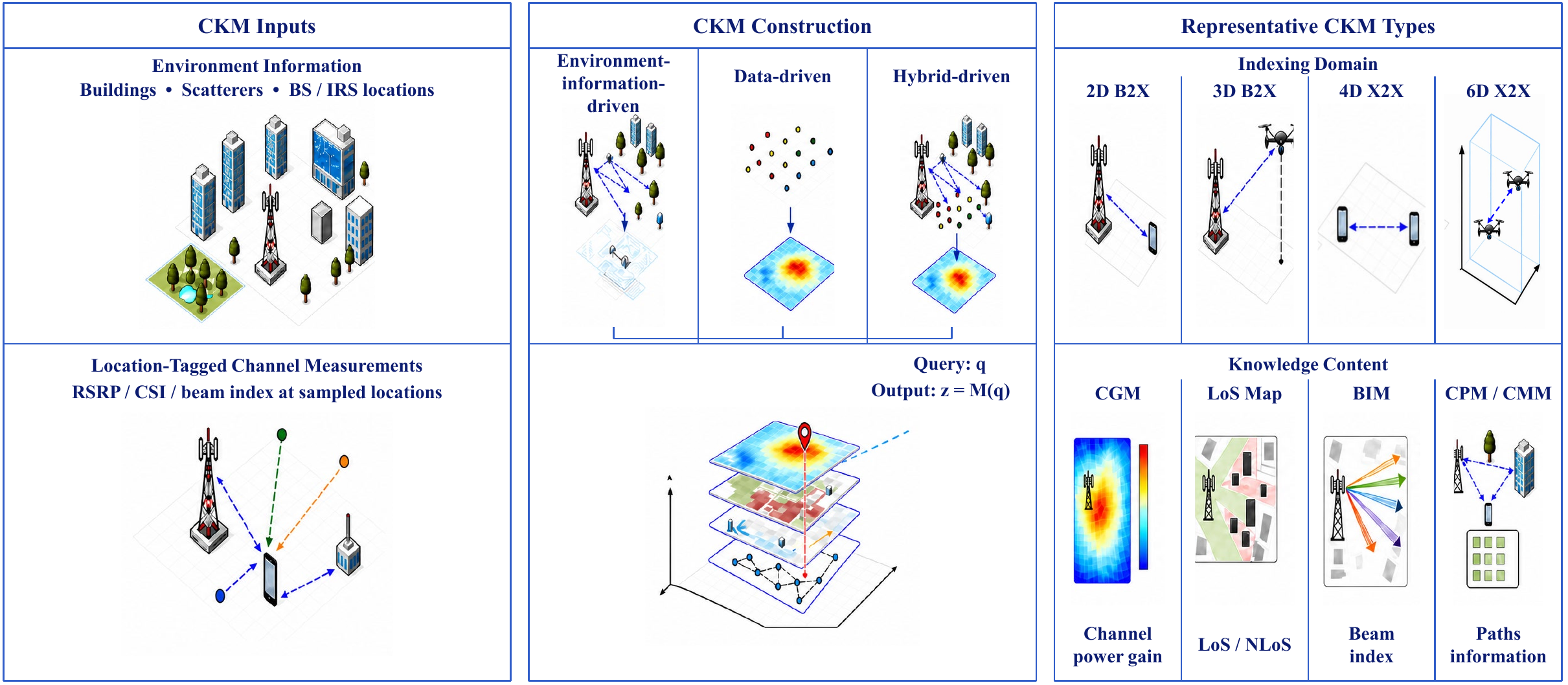}
		\label{Fig_CKM_classification}
	}\\[-1mm]
	\subfloat[]{
		\includegraphics[width=0.98\textwidth]{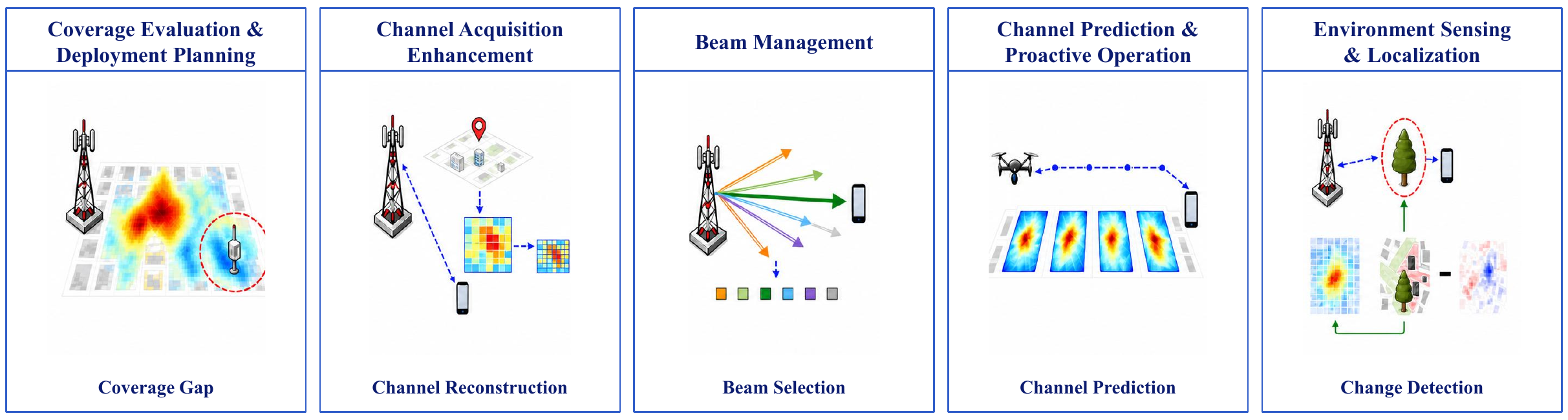}
		\label{Fig_CKM_applications}
	}
	\caption{CKM framework and representative applications: (a) CKM inputs, construction, and representative types; (b) CKM-enabled channel cognition applications.}
	\label{Fig_CKM_overview}
\end{figure*}

Channel cognition enables wireless networks to acquire, organize, infer, and exploit environment-aware channel knowledge across space, time, and frequency domain. Unlike conventional channel acquisition, which is performed reactively after link establishment, CKM-enabled cognition can provide prior channel knowledge for unvisited locations and reduce measurement overhead in dense, mobile, wideband, and large-array systems. As illustrated in Fig.~\ref{Fig_CKM_classification}, environmental information and location-tagged channel measurements can be integrated through environment-information-driven, data-driven, or hybrid-driven construction methods. The resulting CKMs can be indexed in BS-to-any (B2X) or any-to-any (X2X) domains and can represent different forms of channel knowledge, such as channel gains, line-of-sight (LoS) states, recommended beam indices, propagation paths, and channel matrices. This input-construction-indexing-knowledge framework provides the organizational basis for the following review of CKM fundamentals, construction methods, representative applications, and related channel cognition technologies.

\subsection{Channel Cognition and CKM Fundamentals}

Wireless channels fundamentally depend on transceiver locations and the surrounding propagation environment \cite{zeng2021CKM,zeng2024CKMtut}. For a given radio configuration and environment $\mathcal{E}$, a CKM provides a queryable representation of this location-dependent propagation relationship. Let $\mathbf{q}\in\mathcal{Q}\subseteq\mathbb{R}^{D}$ denote an input $D$-dimensional location vector comprising the transmitter and/or receiver positions, and let $\mathbf{z}\in\mathcal{Z}$ denote the channel knowledge of interest. The CKM can be formally represented as
\begin{equation}
	\mathcal{M}_{\mathcal{E}}: \mathcal{Q} \rightarrow \mathcal{Z}, \qquad
	\mathbf{z}=\mathcal{M}_{\mathcal{E}}(\mathbf{q}),
	\label{Eq_CKM_mapping}
\end{equation}
where the output space $\mathcal{Z}$ is application-dependent and may comprise real-valued channel power gains, discrete LoS states or beam indices, path parameters, or complex-valued CSI. The effects of geometry, materials, and radio configuration are captured by $\mathcal{M}_{\mathcal{E}}$ through environmental information, location-tagged measurements, or both. In practice, this mapping may be stored as a location-tagged spatial database or an image-like map, or represented by a learned model \cite{zeng2024CKMtut,ren2026CKMconstruction}. Given the transceiver locations, the CKM can be queried to provide prior channel knowledge, thereby reducing channel uncertainty and enabling operation with little or no online training when complete real-time channel acquisition is costly or infeasible.

According to their spatial indexing domains, CKMs can be broadly classified into B2X CKMs and X2X CKMs. For a B2X CKM, the BS location is fixed, and only the user location is required as the query input. The input dimension is typically $D=2$ for terrestrial users and $D=3$ for users distributed in three-dimensional (3-D) space, such as UAVs. By contrast, an X2X CKM characterizes the channel between an arbitrary transmitter-receiver pair. Because both endpoints may move, the locations of both the transmitter and receiver are required as inputs, resulting in a four-dimensional (4-D) representation for ground-to-ground links and up to a six-dimensional (6-D) representation for fully 3-D links. X2X CKMs provide a more general representation of channel conditions but impose higher requirements on data collection, storage, and model construction \cite{zeng2024CKMtut,ren2026CKMconstruction}.

CKMs can also be classified according to the channel knowledge they provide. At a coarse granularity, a CKM may provide location-specific large-scale channel knowledge, such as LoS conditions, channel gains, dominant path directions, and recommended beam indices. At a finer granularity, it may capture multipath gains, delays, angles of arrival (AoAs), angles of departure (AoDs), or complete CSI. Representative CKM types therefore include the channel gain map (CGM), channel path map (CPM), beam index map (BIM), and channel matrix map (CMM). These maps offer different trade-offs among information richness, localization accuracy, storage overhead, and update complexity. Together, they transform location-tagged channel observations into reusable environment-aware channel knowledge and form the foundation for proactive channel cognition in 6G networks.

\subsection{CKM Construction Methods}

CKM construction aims to obtain the mapping in \eqref{Eq_CKM_mapping} from channel measurements and, when available, physical environment information. As summarized in Table~\ref{Tab_CKM_construction}, existing approaches can be mainly classified as environment-information-driven, data-driven, or hybrid environment- and data-driven methods according to their primary information sources. These three categories are reviewed below.

Environment-information-driven approaches construct CKMs mainly from prior knowledge of the physical propagation environment. Typical inputs include city maps, building layouts, terrain information, object positions, material properties, and radio deployment parameters. Ray tracing is a representative example: it tracks propagation paths between the transmitter and receiver while accounting for reflection, diffraction, and scattering, thereby generating location-specific channel knowledge from detailed environment models \cite{ren2026CKMconstruction}. When the environment model is accurate, these approaches can generate high-fidelity and physically interpretable CKMs while reducing the need for dense field measurements. However, they require accurate geometry and material information, incur high computational complexity in large-scale scenarios, and remain difficult to update in dynamic environments.

Data-driven approaches construct CKMs primarily from location-tagged channel measurements without requiring explicit environment reconstruction. Classical interpolation-based methods estimate channel knowledge at unmeasured locations from sparse measurements. Representative examples include Kriging, kernel regression, and matrix completion \cite{sun2022interpolation,li2022CKMconstruction}. Kriging and kernel regression rely on explicit spatial correlation or kernel assumptions, whereas matrix completion usually imposes a low-rank map prior rather than an explicit electromagnetic propagation model. Learning-based and generative methods further exploit data priors to recover CKMs from incomplete or noisy observations. For example, CKMDiff employs a diffusion model to support denoising, inpainting, and super-resolution \cite{fu2025CKMDiff}. Data-driven approaches are relatively straightforward to deploy when measurements are available and can adapt to site-specific propagation effects. However, their accuracy depends strongly on the density, coverage, and freshness of the measurements. Their ability to extrapolate to unobserved regions also remains limited, particularly when constructing high-dimensional X2X CKMs.

Hybrid environment- and data-driven approaches jointly exploit physical environment information, channel measurements, and propagation priors. Image-based learning methods typically represent a two-dimensional (2-D) CKM as an image-like map and use environmental layouts, transmitter locations, and channel samples to reconstruct channel knowledge. Representative methods include RadioUNet and RadioTransformer \cite{levie2021RadioUNet,li2025RadioTransformer}. RadioUNet uses a convolutional encoder-decoder architecture to capture multi-scale spatial features, while RadioTransformer applies self-attention to model long-range dependencies.

Within the hybrid category, scatterer-centric CGM modeling combines limited channel-gain measurements with a structured multipath propagation prior. Instead of explicitly simulating all electromagnetic interactions as in ray tracing, the scatterer model represents the average channel power over spatial grids using dominant scatterers, path-loss coefficients, and scatterer response coefficients estimated from measurements \cite{sun2025scattererCGM}. A subsequent extension, the 3-D virtual scatterer model, further treats the number and positions of scatterers as tunable parameters and models the angular correlation of their response coefficients. Consequently, a huge physical scatterer may be divided into multiple virtual scatterers, while nearby moderate scatterers with similar effects may be merged as one virtual scatterer \cite{sun2026virtualScattererCGM}. This approach occupies an intermediate position between deterministic environment modeling and pure measurement interpolation and require neither a complete environment model nor dense measurements.

Another hybrid approach consists of physics-aware learning methods that combine sparse measurements with propagation constraints or physically structured representations. These methods extend CKMs beyond discrete 2-D maps. NeRF$^2$ extends neural radiance fields (NeRFs) from the optical domain to the radio-frequency domain by using neural networks to represent complex-valued signals \cite{zhao2023NeRF2}. It combines learned volumetric representations with physical propagation principles to infer signal amplitude and phase at arbitrary locations from sparse measurements, although its volumetric rendering process incurs high training complexity. Wireless radiance fields with Gaussian splatting (WRF-GS) replace implicit volumetric rendering with explicit 3-D Gaussian splatting (3DGS) primitives \cite{wen2025WRFGS}, thereby enabling faster rendering, lower complexity, and higher sample efficiency. Bidirectional wireless Gaussian splatting (BiWGS) further incorporates bidirectional scattering, distance-dependent attenuation, phase rotation, and channel reciprocity to support 6-D X2X CKMs \cite{zhou2025BiWGS}. Although these approaches support accurate, continuous, and high-dimensional CKM construction, they introduce challenges in multi-source data alignment, model training, storage, online updating, and cross-scenario generalization.

\begin{table*}[!t]\scriptsize
	\centering
	\caption{Comparison of representative CKM construction approaches.}
	\label{Tab_CKM_construction}
	\renewcommand{\arraystretch}{1.15}
	\setlength{\tabcolsep}{3pt}
	\begin{tabularx}{\textwidth}{|>{\centering\arraybackslash}p{2.6cm}|p{3.0cm}|p{3.8cm}|X|X|}
		\hline
		\centering\arraybackslash
		\textbf{Category} & \centering\textbf{Main Inputs} & \centering\textbf{Representative Methods} & \centering\textbf{Advantages} & \centering\arraybackslash\textbf{Limitations} \\ \hline
		Environment-information-driven & Environmental maps, geometry, materials, and radio deployment parameters & Ray tracing; deterministic propagation simulation; environment-aware propagation modeling & Physically interpretable; high fidelity when environment information is accurate; less dependent on dense field measurements & Requires accurate environment modeling; high computational complexity; difficult to update online in dynamic environments \\ \hline
		Data-driven & Location-tagged channel measurements and spatial sampling records & Kriging; kernel regression; matrix completion; generative map recovery, e.g., CKMDiff & No detailed environment model required; adapts to site-specific measurements; effective for interpolation, completion, and super-resolution & Sensitive to measurement density, coverage, and freshness; limited extrapolation and cross-scenario generalization; challenging for high-dimensional CKMs \\ \hline
		Hybrid environment- and data-driven & Environmental information, channel measurements, and physical or structural priors & RadioUNet; RadioTransformer; scatterer model; virtual scatterer model; NeRF$^2$; WRF-GS; BiWGS & Fuses complementary information sources; supports rich 3-D or continuous representations; suitable for complex and high-dimensional scenarios & Requires multi-source data alignment; high training, storage, and updating costs; robustness and generalization remain challenging \\ \hline
	\end{tabularx}
\end{table*}

In summary, CKM construction has evolved from approaches that rely solely on explicit environment information or channel measurements toward hybrid-driven frameworks that integrate environmental knowledge, measured data, and propagation priors. Despite these advances, accurate, real-time, and scalable CKM construction in dynamic environments remains an important topic for future study.

\subsection{CKM-Enabled Channel Cognition Applications}

As illustrated in Fig.~\ref{Fig_CKM_applications}, once constructed, a CKM serves as a reusable spatial knowledge base that converts location-tagged channel observations into actionable channel knowledge. It supports coverage evaluation and deployment planning, provides link-specific priors for channel acquisition, narrows the candidate beam set, enables channel prediction along future trajectories, and facilitates the detection or localization of environmental changes. These capabilities also provide the channel and environmental priors required for subsequent proactive channel reconfiguration.

\subsubsection{Coverage Evaluation and Deployment Planning}
Conventional coverage analysis often relies on extensive site surveys or simplified propagation models. By contrast, a CKM provides location-specific channel gains, LoS conditions, and interference information across the target area. Such knowledge enables the network to identify coverage holes and evaluate how infrastructure elements such as BSs, APs, relays, and IRSs affect the propagation environment before actual deployment or transmission \cite{zeng2024CKMtut}. CKM-based coverage evaluation can therefore support infrastructure deployment, coverage enhancement, and reconfiguration planning.

\subsubsection{Channel Acquisition Enhancement}
In large-scale antenna array systems, exhaustive channel estimation consumes substantial time-frequency resources. A channel matrix map or channel path map can provide coarse channel estimates directly from transceiver locations. When online measurements remain necessary, CKM priors reduce the number of unknown parameters and support channel acquisition with limited online training and substantially lower pilot overhead \cite{wu2024CKMbeamforming,zeng2024CKMtut}.

\subsubsection{Beam Management}
In high-frequency and large-array systems, conventional beam sweeping searches over a large number of candidate beam pairs and thus incurs high training latency. A channel angle map or beam index map can instead provide dominant propagation directions or recommend a small set of candidate beams based on transceiver locations. Beam training can then be confined to these candidates, significantly reducing overhead while maintaining reliable alignment \cite{wu2024CKMbeamforming,zeng2024CKMtut}.

\subsubsection{Channel Prediction and Proactive Network Operation}
By combining a CKM with predicted user trajectories, the network can estimate channel conditions at locations that users have not yet reached and make decisions in advance. For example, a UAV can anticipate coverage holes, severe blockage, and strong interference along a future route rather than reacting only after the channel quality has degraded. The resulting predictive channel information can support scheduling, handover, power control, node activation, and the triggering of reconfiguration \cite{zhang2021radiomapUAV,zeng2023CKMpredictive}.

\subsubsection{Environment Sensing and Localization}
Because a CKM contains location-tagged channel features, its mapping can be used inversely to infer the positions of users, obstacles, and scatterers. For instance, LoS maps can improve anchor selection in non-line-of-sight (NLoS) environments, while channel gain, angle, and path maps can serve as robust fingerprints for localization. Deviations between real-time measurements and the stored CKM can also indicate environmental changes, thereby enabling the detection and tracking of dynamic objects. The resulting sensing information can in turn be used to update the CKM and improve subsequent network decisions \cite{long2022CKMlocalization,zeng2024CKMtut}.

Together, these applications enable the network to identify where and why unfavorable propagation occurs and to select appropriate reconfiguration actions, such as antenna repositioning or rotation, IRS deployment and configuration, and the selection of candidate propagation paths. Post-reconfiguration measurements can then be used to update the CKM, thereby forming the cognition-reconfiguration-feedback loop. This loop connects CKM-based cognition to the MA- and IRS-based reconfiguration techniques reviewed in the following sections.

\subsection{Related Channel Cognition Technologies}

Although CKM is a prominent enabler of channel cognition in 6G networks, the broader channel cognition paradigm encompasses multiple complementary technologies. These technologies acquire, represent, infer, and exploit channel-related knowledge from the perspectives of spectrum awareness, spatial radio-frequency (RF) characterization, propagation modeling, temporal channel forecasting, system-level virtualization, and AI-based channel learning. The following discussion briefly reviews these technologies and clarifies their respective roles in channel cognition.

Cognitive radio is one of the earliest environment-aware wireless paradigms. It enables terminals and networks to sense spectrum usage, detect interference, and adapt access strategies accordingly \cite{haykin2005cognitive}. Within the broader context of channel cognition, cognitive radio primarily provides spectrum-domain knowledge, including awareness of spectrum occupancy, interference conditions, and dynamic access opportunities.

Radio maps store spatially indexed RF observations, such as received signal strength, interference level, spectrum occupancy, and aggregate coverage, and support coverage analysis, localization, and resource management \cite{bi2019radiomap}. In many practical radio maps, the recorded value jointly depends on transmit power, the superposition of signals from multiple transmitters, interference, and propagation loss. It therefore reflects the overall RF service condition rather than isolating a specific channel quantity between a particular transmitter-receiver pair. Consequently, radio maps usually provide coarser spatial awareness than maps that explicitly index well-defined channel features, although they remain effective for identifying coverage holes, interference hot spots, and service-quality variations \cite{zeng2024CKMtut}.

Geometry-based and statistical channel models provide mechanism-level knowledge of wireless propagation. Standardized channel models, including those specified in the Third Generation Partnership Project (3GPP) TR~38.901, together with analytical propagation models, characterize path loss, shadowing, clustering, angular spread, and delay spread \cite{3gpp2019studyo,rappaport2019wireless}. They explain how propagation mechanisms and statistical regularities shape channel behavior and provide tractable tools for analysis, simulation, and system evaluation.

Channel prediction addresses the temporal dimension by forecasting future CSI or communication performance from historical observations, user mobility, and environmental dynamics \cite{XJ_CE,liu2024llm4cp}. This capability is particularly important for mobile users, time-varying interference, and low-latency proactive control, where future channel states must be anticipated before conventional estimation becomes available.

Digital twin technology provides system-level cognition by creating a virtual representation of a physical network for monitoring, simulation, testing, and optimization \cite{khan2022digitaltwin}. In wireless communications, a digital twin may integrate traffic models, protocol stacks, network topology, user mobility, and propagation environments, thereby enabling network policies to be evaluated before they are applied to the physical system.

More recently, AI and foundation models have expanded channel cognition by learning complex mappings among environments, network configurations, and channel responses \cite{yu2024channelgpt,liu2024llm4cp}. For example, ChannelGPT generates channel-related parameters and map information from multimodal environmental data, while large language model for competitive programming (LLM4CP) adapts large language models (LLMs) to predict future CSI sequences. These methods support multi-source information fusion, missing-data completion, temporal prediction, and knowledge transfer across scenarios.

In summary, cognitive radio, radio maps, channel models, channel prediction, digital twins, and channel-oriented AI models provide spectrum-domain, spatial, mechanism-level, temporal, system-level, and learning-based knowledge, respectively. Within this broader landscape, CKM occupies a specific but complementary position by organizing location-dependent channel features into a queryable spatial representation. Information and models provided by the other technologies can support CKM construction and updating, while CKMs can provide spatial priors for prediction, simulation, and network decision-making. Together, these technologies form a multidimensional foundation for proactive channel cognition in future 6G networks.

\section{Transceiver-Side Channel Reconfiguration through MA}

Building upon channel cognition, channel reconfiguration further improves wireless systems through adjusting transceiver structures and/or programming signal propagation environments. At the transceiver side, intelligent antenna technologies, such as MAs \cite{010,shao2024Mag6DMA}, reconfigurable antennas \cite{026}, fluid antennas \cite{027}, pinching antennas \cite{liu2025pinching}, and rotatable antennas \cite{zheng2025rotatable}, can exploit additional spatial and electromagnetic DoFs by adjusting antenna positions, orientations, polarizations, radiation characteristics, or array structures. At the environment side, IRSs and their advanced variants, including simultaneously transmitting and reflecting surfaces \cite{10133841}, beyond-diagonal IRS/RIS \cite{li2022beyond}, and stacked intelligent metasurfaces \cite{an2024stacked}, can reshape signal propagation through controllable reflection and/or transmission of metasurfaces in the environments. This section focuses on transceiver-side channel reconfiguration enabled by MAs. By exploiting continuous translational and rotational DoFs, MAs can improve the effective channel conditions. We first introduce the hardware architectures and performance advantages of MAs, followed by a discussion of key design issues and other relevant technologies.

\subsection{Architecture}

\begin{figure*}[!t]
	\centering
	\includegraphics[width=150mm]{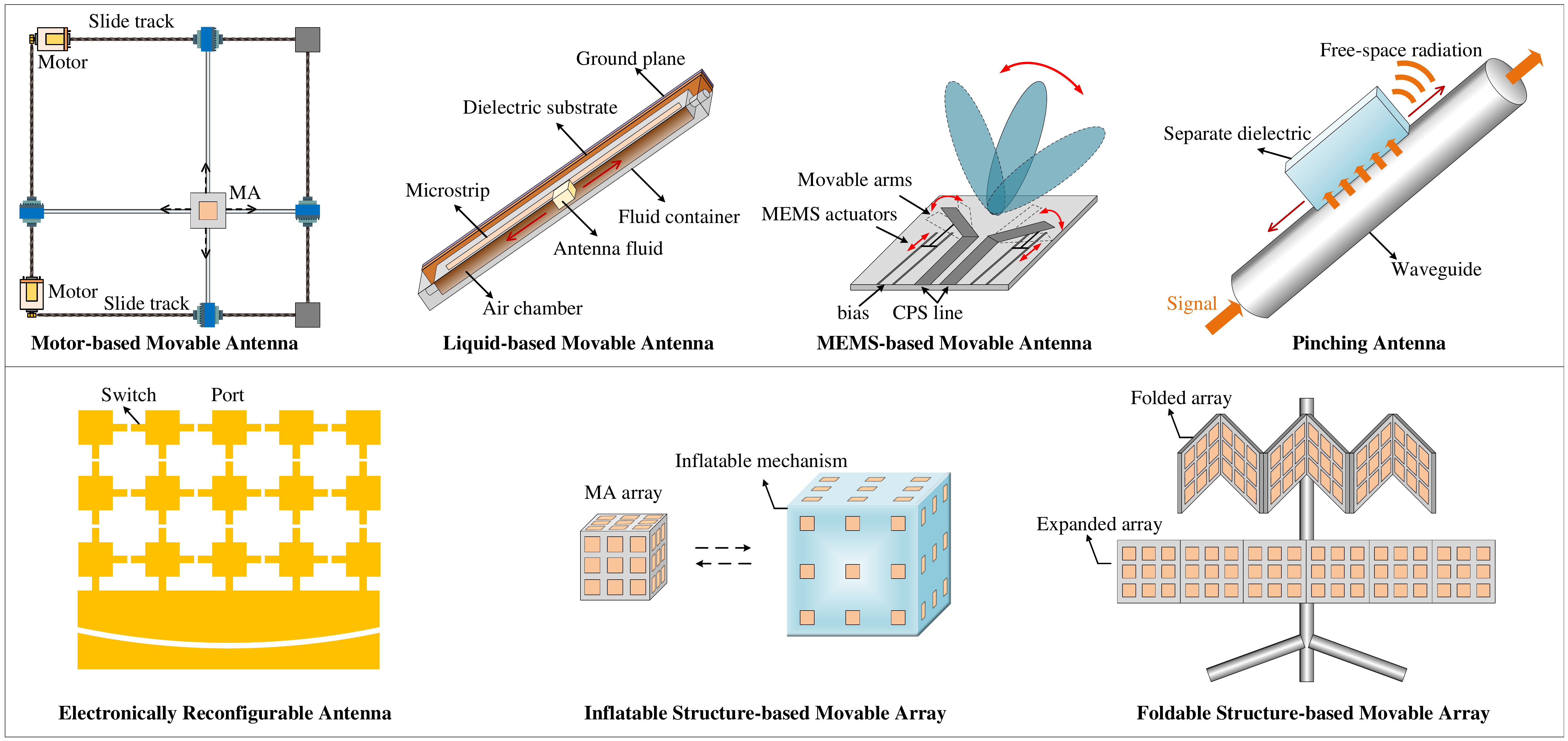}
	\caption{Typical implementation methods for antenna movement and reconfiguration.}
	\label{Fig_Implementations}
\end{figure*}

The practical realization of MA systems is determined by their underlying hardware architectures \cite{zhu2023MAMag,zhu2022MAmodel,ma2022MAmimo,shao2024Mag6DMA, shao20246DMA,shao2025tutorial}. As shown in Fig.~\ref{Fig_Implementations}, different implementation architectures provide distinct DoFs and employ various actuation mechanisms to achieve these DoFs.

\subsubsection{Classification}

The spatial adaptability of MA systems is primarily characterized by two different DoFs, i.e., position and orientation.

\textit{Position}: This DoF enables the relocation of the antenna phase center within a one-dimensional (1-D), 2-D, or 3-D spatial region \cite{zhu2023MAMag,zhu2022MAmodel}. By dynamically repositioning the radiating element subarray on the wave-length order, the system can exploit favorable channel conditions through adjusting the phase responses of different channel paths. For large-scale positioning, the LoS conditions, shadowing, and path loss between the transceivers can be proactively reconfigured \cite{fu2025extremely,lu2026wireless,ding2024flexible}.

\textit{Orientation}: In addition to spatial positioning, orientation provides rotational DoFs, including pitch, yaw, and roll \cite{shao2024Mag6DMA, shao20246DMA,shao2025tutorial,zheng2025rotatable,zhengtian2025rotatable}. Optimizing the antenna orientation is particularly important for directional antennas, as it enables accurate alignment of the main radiation beam toward target transceivers and facilitates polarization matching with incident electromagnetic waves. Combining both the 3-D position and 3-D orientation movement, the general form of six-dimensional MA (6DMA) system can offer the highest DoF for spatial channel reconfiguration at the transceiver \cite{shao2024Mag6DMA, shao20246DMA,shao2025tutorial}.

\subsubsection{Implementation Methods}

\begin{table*}[!t]
	\scriptsize
	\centering
	\caption{Performance comparison of representative antenna movement implementation methods.}
	\label{tab:Key-Characteristics-of-MA}
	\renewcommand{\arraystretch}{1.15}
	\setlength{\tabcolsep}{3pt}
	\begin{tabularx}{\textwidth}{|>{\centering\arraybackslash}p{4.8cm}|
			>{\centering\arraybackslash}p{4.8cm}|
			>{\centering\arraybackslash}p{3.5cm}|
			>{\centering\arraybackslash}X|}
		\hline
		\textbf{Implementation Methods} &
		\textbf{Tuning Range} &
		\textbf{Tuning Speed} &
		\textbf{Power Consumption} \\ \hline
		
		Motor &
		Large &
		Moderate (ms$\sim$s) &
		High \\ \hline
		
		Liquid &
		Constrained by linear fluidic channels &
		Moderate (ms$\sim$s) &
		Moderate \\ \hline
		
		MEMS &
		Small &
		Fast (µs$\sim$ms) &
		Low \\ \hline
		
		Pinching Antenna &
		Constrained by linear waveguides &
		Moderate (ms$\sim$s) &
		Moderate \\ \hline
		
		Electronically Reconfigurable Antenna &
		Small &
		Fast (ns$\sim$µs) &
		Low \\ \hline
		
		Inflatable Structure &
		Fixed (Stowed-to-deployed) &
		Slow (s) &
		High \\ \hline
		
		Foldable Structure &
		Fixed (Stowed-to-deployed) &
		Slow (s) &
		High \\ \hline
		
	\end{tabularx}
\end{table*}

As shown in Fig.~\ref{Fig_Implementations}, depending on the required movement scale, ranging from micro-scale element tuning to macro-scale array deployment, various mechanical and electronic approaches can be adopted to achieve these DoFs.

\textit{Motor-based Methods}: Conventional mechanical architectures employ external actuators, such as precision gears, linear tracks, and stepper motors, to physically translate or rotate the antenna \cite{zhu2023MAMag}. These systems provide large tuning ranges and strong load-bearing capability, but generally suffer from high power consumption and heavy hardware structures.

\textit{Liquid-based Methods}: In these architectures, conductive or dielectric fluids, such as liquid metals, are displaced within confined microfluidic channels. Fluid motion can be driven by nano-pumps, syringes, or electrowetting techniques, enabling flexible reconfiguration of the effective radiating structure while remaining constrained by the predefined channel geometry.

\textit{Micro-Electromechanical Systems (MEMS)-based Methods}: MEMS-based approaches utilize miniaturized electromechanical components to achieve micro-scale positional adjustments \cite{zhu2023MAMag}. Such systems offer ultra-fast response times, typically in the microsecond range, together with low power consumption, making them suitable for fine-grained antenna tuning.

\textit{Pinching Antenna-based Methods}: Pinching antennas generate a repositioning radiating element along a fixed dielectric waveguide through mechanical or electrical tuning \cite{ding2024flexible}. Mechanical tuning physically moves a secondary dielectric structure near the waveguide to create the pinching effect, whereas electrical tuning employs integrated active components (e.g., PIN diodes or varactors) to achieve rapid and reliable position control.

\textit{Electronically Reconfigurable Antenna-based Methods}: To overcome the latency and mechanical wear associated with physical actuation, electronically controlled mechanisms, such as RF switching networks and reconfigurable metamaterials, can emulate mechanical translation or rotation \cite{chen2025remaa}. These approaches provide ultra-fast response times and superior platform integration, although their effective tuning range is generally more limited than that of physical movement.

\textit{Inflatable Structure-based Methods}: Inflatable architectures are mainly adopted for large-scale deployable arrays in scenarios with stringent stowage constraints, such as aerospace and satellite applications. These structures utilize internal gas pressure to expand compactly stored radiating surfaces into large operational apertures.

\textit{Foldable Structure-based Methods}: Foldable architectures exploit origami-inspired hinges and modular linkages to transition between compact and deployed configurations. Such structures are particularly suitable for terrestrial BSs, where retractable array geometries can reduce environmental impacts, such as wind loading, when full aperture deployment is unnecessary. A performance summary table is given in Table~\ref{tab:Key-Characteristics-of-MA} to facilitate a clear comparison of the trade-offs associated with different implementation methods.

\subsection{Performance Advantages}
\begin{figure*}[!t]
	\centering
	\includegraphics[width=150mm]{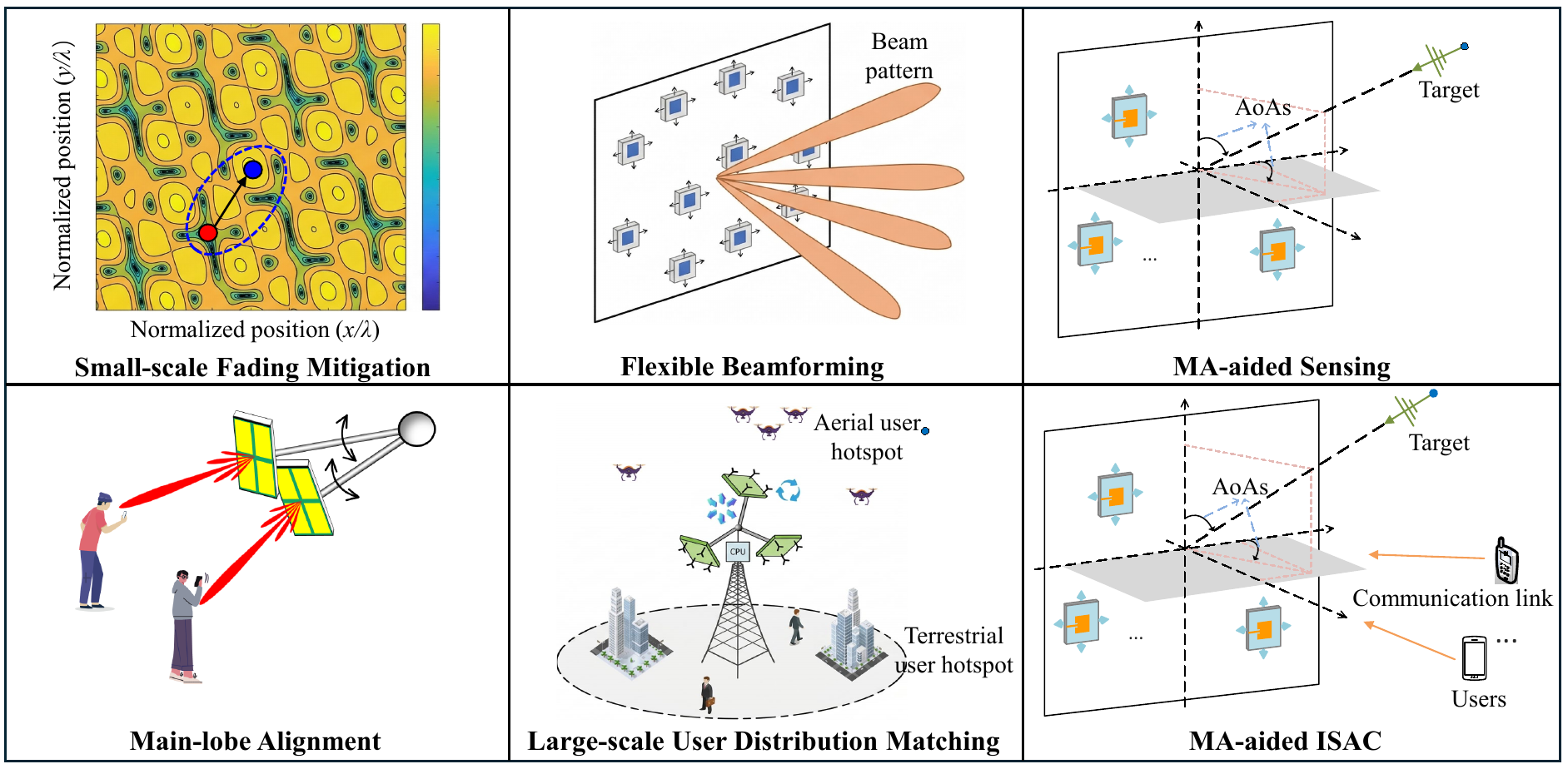}
	\caption{Performance advantages of MA.}
	\label{Fig_MA_advantage}
\end{figure*}

The MA-enabled wireless systems fundamentally transforms how wireless transceivers interact with electromagnetic environments. As illustrated in Fig.~\ref{Fig_MA_advantage}, MA can significantly benefit both communication and sensing paradigms.

\subsubsection{Small-scale Fading Utilization}
The MA system can exploit the small-scale fading of wireless channels. Instead of suffering through severe signal degradation in deep fading regions, MAs can dynamically relocate to favorable spatial regions where multipath components are constructively superimposed. This localized adjustment effectively improves the equivalent channel gain and maximizes the received signal-to-noise ratio (SNR) \cite{zhu2022MAmodel}. Even in wideband scenarios characterized by frequency-selective fading, antenna position optimization can yield substantial average power gains across the entire operational bandwidth \cite{zhu2024wideband}. Moreover, MAs can exploit the spatial variation of interference by relocating to positions with weaker interference, thereby mitigating interference and improving the desired signal-to-interference-plus-noise ratio (SINR).

\subsubsection{Flexible Beamforming}
In multi-antenna systems, the MA array can form flexible beam patterns via antenna position optimization \cite{zhu2023MAarray, ma2024multi, wang2024flexible}. The joint optimization of the array geometry and the antenna weights facilitates the generation of customized multi-beams or wide-beam coverage, surpassing the conventional systems that can only optimize the antenna weights. Furthermore, MAs can steer high-gain main-lobes toward intended directions while strategically placing deep nulls in the directions of interferers. In multi-user setups, this flexibility directly reduces spatial channel correlation, thereby suppressing inter-user interference and enhancing spatial multiplexing \cite{zhu2023MAmultiuser, xiao2023multiuser}.

\subsubsection{Main-lobe Alignment}
Beyond translational movement, rotational DoFs offer a complementary approach for main-lobe alignment, particularly for highly directional antennas \cite{shao2024Mag6DMA, shao20246DMA,shao2025tutorial,zheng2025rotatable,zhengtian2025rotatable}. By mounting the MA on a rotatable mechanism, the system can adjust its physical orientation in 3-D space. This joint azimuth and elevation rotation allows the broadside of the array to maintain precise alignment with the dominant AoA or AoD. Consequently, main-lobe alignment maximizes the effective directional gain and eliminates misalignment losses without requiring complex and power-intensive phase shifter reconfigurations for extreme steering angles.

\subsubsection{Large-scale User Distribution Matching}
For macro-level network adaptations, slow-moving or structurally flexible architectures (such as 6DMAs or extremely large-scale MAs (XL-MAs)) can actively align their macro-geometry with dynamic network demands according to varying user distributions \cite{fu2025extremely, yan2025movable, jiang2026statis6dma, yan2026slowMA}. By relocating antenna subarrays close to or reorienting massive antenna facets toward localized hotspot areas (e.g., clustered UAV networks or dense terrestrial users), the system can achieve improved coverage performance. This macroscopic adaptation operates on a slower timescale, matching statistical spatial user distributions rather than instantaneous small-scale fading.

\begin{figure}[!t]
	\centering
	\includegraphics[width=80mm]{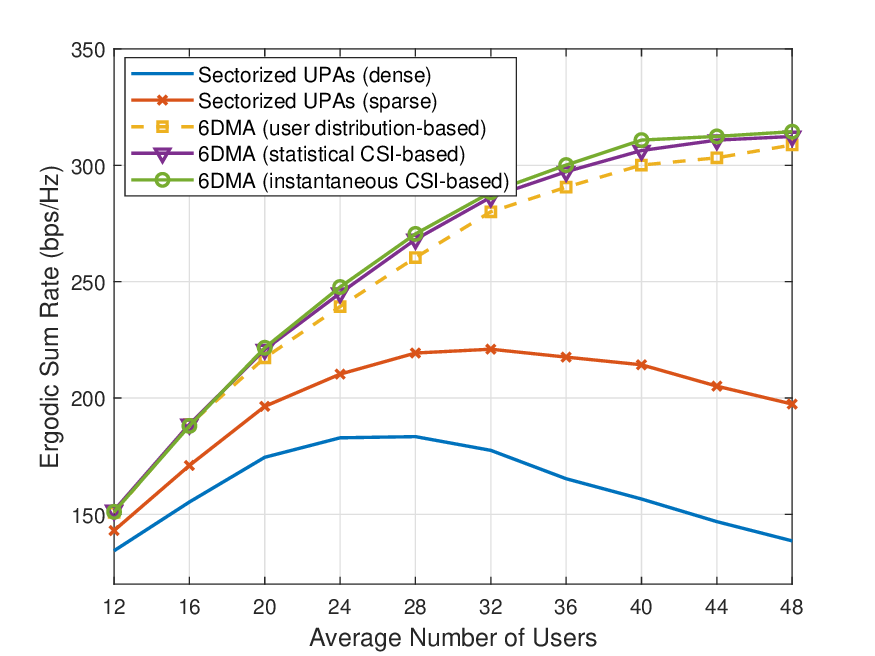}
	\caption{Performance comparison of 6DMA arrays optimized over different timescales and conventional three-sector FPA arrays under different averaged number of users.}
	\label{Fig_6dMA_comparison}
\end{figure}

\subsubsection{Sensing}
In wireless sensing, the antenna movement can enlarge the effective virtual aperture, which thus improves the spatial resolution \cite{ma2024MAsensing}. For far-field target localization, optimizing the continuous antenna's trajectory or the discrete antennas' positions significantly reduces the Cramér-Rao bound (CRB) and the mean square error (MSE) for AoA estimation. In near-field scenarios, intelligent sampling of spherical wavefronts by dynamically positioned antennas enhances joint range and angle resolution. Unlike traditional synthetic aperture radar (SAR) operating on predefined tracks, MAs enable adaptive and real-time trajectory optimization tailored to the target's state, and the antennas' positions can be optimized to reduce grating lobes and minimize the ambiguity error \cite{ma2025movabletra}.

\subsubsection{ISAC}
Since the MA can improve the sensing and communication performance, it constitutes a highly synergistic hardware foundation for ISAC networks \cite{ma2025MAISAC, lyu2024flexibleISAC}. Specifically, MA provides a unique physical mechanism to balance the distinct requirements of sensing and communication tasks. An MA-aided ISAC BS can strategically distribute its antenna elements to synthesize a sparse, high-resolution aperture for radar sensing, while simultaneously ensuring the antennas' positions minimize inter-user interference for concurrent data transmission. This dynamic geometric adaptability allows the system to seamlessly shift its priority between communication throughput and sensing accuracy based on real-time environmental demands.

Fig.~\ref{Fig_6dMA_comparison} compares the ergodic sum rates of downlink transmission between 6DMA arrays and conventional three-sector fixed-position antenna (FPA) arrays. The simulation setup is detailed in~\cite{yan2026slowMA}, with the BS placed at the center of the cell instead. The carrier frequency is $3$ GHz and $12$ 6DMA subarrays are employed, each composed of a $2\times 2$ antenna array. For the benchmark sectorized FPA systems, three sectors are deployed, each employing a $4\times4$ uniform planar array (UPA). The inter-element spacing is set to half a wavelength for the dense UPA scheme and one wavelength for the sparse UPA scheme. For the 6DMA system, three antenna optimization strategies operating at different timescales are considered: i) user distribution-based design, where the 6DMAs are optimized according to the spatial distribution of users; ii) statistical CSI-based design, where the 6DMAs are optimized based on statistical CSI; and iii) instantaneous CSI-based design, where the 6DMAs are optimized for each instantaneous channel realization. As shown in Fig.~\ref{Fig_6dMA_comparison}, all 6DMA schemes achieve substantially higher ergodic sum rates than their FPA counterparts. Notably, the statistical CSI-based and user distribution-based designs attain performance close to that of the instantaneous CSI-based design. These results demonstrate significant performance gains of 6DMA systems over FPA systems, while indicating that adaptation to statistical CSI, and even user distribution, may be sufficient to capture the majority of these gains without requiring real-time reconfiguration based on instantaneous CSI.

\subsection{Design Issues}
While MAs offer significant performance gains for wireless communication and sensing, their practical deployment raises new design challenges beyond those encountered in conventional FPA systems. This subsection reviews the key design issues of MA-enabled wireless systems, including channel modeling, channel acquisition, antenna movement optimization and scheduling, as well as prototype development and experimental demonstrations.

\subsubsection{Channel Modeling}

Accurate channel modeling is essential for the design and performance evaluation of MA systems. Channel cognition (e.g., CKM) provides the underlying environmental knowledge and predictive information for channel modeling. By leveraging environment-aware and location-specific channel knowledge at different spatial granularities, MA channel models can accurately characterize and parameterize the continuous variations in wireless channels induced by antenna movement. Existing MA channel models can be broadly categorized into those based on physical path propagation and those based on statistical channel properties.

The physical field-response channel model provides a rigorous framework that directly relates the wireless channel to the underlying propagation environment \cite{zhu2022MAmodel, ma2022MAmimo, zhu2024nearfield, zhu2024wideband}. By expressing the channel as an explicit function of the transceivers' antenna position and orientation, this model characterizes the wireless channel as the superposition of multiple propagation paths, each associated with a complex path gain, AoA, and AoD. Although initially developed for deterministic propagation environments, recent extensions have incorporated statistical channel modeling to characterize random channel variations caused by factors like the local movement of users \cite{yan2025movable}. Furthermore, the field-response channel model exhibits significant flexibility. In wideband systems, distinct field response vectors (FRVs) and path parameters can be assigned to different delay taps \cite{zhu2024wideband}. In near-field scenarios, the conventional plane-wave assumption is replaced by spherical-wave propagation models to accurately characterize the wavefront \cite{zhu2024nearfield}. The main challenge of this model lies in the substantial analytical and computational complexity, which increases significantly with the number of resolvable multipath components.

In contrast, spatial correlation-based statistical models offer an alternative approach \cite{wong2020fluid}. When the underlying statistical assumptions are valid, these models exhibit strong robustness against moderate environmental mismatches. Classical frameworks, such as extensions of Jake’s model, have been adapted to MA systems operating in rich-scattering environments. These models describe channel evolution during antenna movement through spatial-temporal correlation functions combined with fading distributions, such as Rayleigh and Rician fading. Despite their mathematical simplicity, statistical models generally fail to capture fine-grained multipath geometries and antenna-specific radiation characteristics. Consequently, they are more appropriately regarded as macroscopic approximations of the more detailed field-response models, particularly suitable for environments with densely distributed random scatterers \cite{zhu2022MAmodel}.

\subsubsection{Channel Acquisition}

Fully exploiting the spatial DoFs offered by MA systems requires accurate CSI throughout the entire antenna movement region at both the transmitter and receiver \cite{ma2023MAestimation, xiao2023channel}. In this context, channel cognition provides prior information for channel acquisition. Specifically, by leveraging location-specific channel knowledge, it reduces channel uncertainty and narrows the search space of channel parameters for real-time local CSI estimation, thereby enabling high-accuracy channel acquisition with reduced training overhead. The construction of such spatial channel mapping fundamentally depends on the underlying channel model. In particular, model-based acquisition techniques adopt field-response models to estimate physical propagation parameters, whereas model-free approaches are preferable when statistical channel models are adopted or deterministic channel structures are unavailable.

Model-based channel acquisition aims to recover the field response information (FRI), including the path response matrix (PRM) and wave vectors associated with AoAs and AoDs, from a limited set of spatial measurements. Owing to the angular sparsity typically exhibited by dominant propagation paths \cite{ma2023MAestimation, xiao2023channel, zhang2024TensorCE,xiao2024channelwide}, compressed sensing techniques, such as orthogonal matching pursuit (OMP), can reconstruct the complete FRI from a substantially reduced number of measurement samples \cite{wei2025super}. Tensor decomposition methods provide another model-based solution \cite{zhang2024TensorCE}. By arranging MA sampling locations into a UPA structure, the received pilot observations can be represented as a multidimensional tensor. Subsequent tensor decomposition, such as canonical polyadic decomposition, enables gridless super-resolution estimation of wave vectors, which is particularly effective in high-SNR scenarios. More recently, learning-aided model-based approaches have jointly optimized the MA sampling positions and FRI estimation accuracy \cite{jang2025new, jang2026deep}. Another recent approach adopts compressed sensing as initial estimation, with a further denoising neural network to improve FRI estimation accuracy in multiuser wideband MA systems \cite{feng2026deep}. The principal advantage of model-based approaches is their low sampling overhead, which scales primarily with the number of channel paths rather than the physical size of the movement region. However, these methods are generally sensitive to measurement noise and rely heavily on the accuracy of the assumed channel model.

In contrast, model-free channel acquisition avoids imposing explicit structural assumptions on the propagation environment and instead relies on direct spatial sampling followed by interpolation or prediction to estimate unobserved CSI. Simple implementations estimate the CSI at an unknown position using the nearest sampled location \cite{Skouroumounis2023fluidCE}. More advanced probabilistic approaches, such as Bayesian linear regression, model the spatial channel distribution as a Gaussian random field and iteratively refine the posterior channel estimate as additional measurements are collected \cite{zhang2023successive}. In addition, data-driven machine learning techniques employ neural networks trained on discrete channel measurements to predict CSI at arbitrary positions \cite{huang2025cnn}. Although model-free methods are generally more robust to modeling inaccuracies and highly effective in small movement regions, they require dense spatial sampling with sub-wavelength resolution. As a result, the overhead increases with both the dimensionality and size of the antenna movement region.

\subsubsection{Antenna Movement Optimization and Scheduling}

The performance gains achievable by MA systems fundamentally depend on the antenna movement strategy, determining when and where the antennas should be repositioned based on the acquired CSI while accounting for the associated mechanical and temporal costs. CSI-driven movement optimization aims to dynamically adjust antennas' positions and orientations to maximize communication performance metrics, such as SNR \cite{zhu2022MAmodel}, system capacity \cite{ma2022MAmimo}, flexible beamforming capability, and interference suppression \cite{zhu2023MAarray, ma2024multi, wang2024flexible}. Since the relationship between antennas' positions and communication performance is highly nonlinear, the resulting optimization problems are generally non-convex. Existing studies have addressed these challenges using various optimization methodologies, including gradient-based local search algorithms \cite{zhu2023MAmultiuser, zhu2024wideband}, successive convex approximation (SCA) techniques \cite{ma2022MAmimo, ma2024multi}, and global algorithms such as particle swarm optimization \cite{xiao2023multiuser} and graph-theoretic methods based on discretized movement grids \cite{mei2024movable}. More recently, AI-based techniques have demonstrated significant potential in learning efficient movement policies under limited or imperfect CSI conditions, thereby offering an alternative to conventional model-driven optimization frameworks \cite{zhu2023MAMag}.

The appropriate movement timescale largely depends on whether instantaneous or statistical CSI is employed. Instantaneous CSI-based optimization \cite{ma2022MAmimo} is particularly suitable for quasi-static environments with relatively long channel coherence times, where antenna repositioning can effectively adapt to real-time channel variations while maintaining acceptable mechanical and computational overhead. However, in fast-varying wireless environments, continuously tracking instantaneous CSI and frequently repositioning antennas can incur excessive latency and energy consumption. In such cases, optimization based on statistical CSI becomes more practical and efficient. By optimizing long-term performance metrics, such as ergodic capacity, using slowly varying channel statistics including spatial correlation structures and AoA/AoD distributions, the frequency of physical antenna reconfiguration can be substantially reduced while still exploiting the spatial DoFs enabled by MA systems.

Moreover, practical hardware constraints fundamentally limit the achievable performance gains of MA optimization. The system design should explicitly account for actuator speed, which determines the antenna movement latency, as well as the energy consumption associated with antenna movement, particularly in battery-powered devices. In addition, the physically accessible movement space, including translational and rotational constraints in the 1-D, 2-D, or 3-D space, further restricts the feasible optimization region. Therefore, practical MA deployment requires well-coordinated hardware architectures and intelligent scheduling mechanisms to balance communication performance gains against movement delay, energy expenditure, and implementation complexity.

\subsubsection{Prototypes and Experiments}

Transitioning MA systems from theoretical concepts to practical deployment requires hardware prototypes and experiments. Recently, the performance gains of MA-enabled communications have been experimentally validated through various prototype implementations \cite{dong2024movable, wang2025channelFAS, shen2024designv2, zhang2024pixelv2}. For example, a mechanically actuated MA prototype achieved ultra-precise position control with a movement resolution of 0.05 mm. Field measurements demonstrated substantial spatial diversity gains, where moving the antenna over a $6\lambda$ horizontal trajectory at 3.5 GHz produced received signal power variations exceeding 40 dB \cite{dong2024movable}. Moreover, MA systems have been shown to outperform FPA systems in mitigating multipath fading, improving spectral efficiency by up to 10\% at 300 GHz \cite{wang2025channelFAS}. An MA prototype supporting 3-D position adjustment together with 1-D orientation control showed the received power variations exceeding 60 dB \cite{ma2026survey}. Besides motor-driven implementations, alternative architectures based on fluid and pixel antennas have also been demonstrated. In multi-user scenarios, dual-channel fluid antennas reduced the outage probability by 57\% and increased the multiplexing gain to 2.27 \cite{shen2024designv2}, while pixel antennas achieved approximately 30 dB power variation at 2.5 GHz, validating their capability for spatial multiple access \cite{zhang2024pixelv2}.

These prototype implementations highlight several important design challenges. In particular, the achievable performance gains are highly dependent on accurate path state information (PSI) estimation \cite{dong2024movable}. Consequently, future MA prototypes should adopt a joint hardware-algorithm design that tightly integrates high-speed and high-precision actuation with real-time PSI acquisition and tracking, enabling adaptive antenna reconfiguration and fully realizing the theoretical advantages of MAs in dynamic wireless environments.

\subsection{Other Relevant Technologies}

Based on the concept of antenna movement and reconfiguration, several advanced antenna technologies have also been investigated to exploit additional design flexibility of wireless systems. For example, reconfigurable antennas can dynamically adjust their operating frequency, polarization, and radiation pattern by changing their electrical or physical structures \cite{costantine2015reconfigurable}. These capabilities are important for multi-band communication, interference suppression, and system capacity improvement without significantly increasing the antenna size.  Fluid antennas use conductive liquids, such as liquid metals or ionized solutions, inside dielectric capillaries to dynamically change the antenna shape and position \cite{wong2020fluid}. This enables the system to exploit spatial diversity and improve communication performance in severe fading environments. Rotatable antennas achieve beam steering through mechanical rotation \cite{zheng2025rotatable,zhengtian2025rotatable}, while pinching antennas adjust the localized radiation point along a fixed dielectric waveguide \cite{ding2024flexible}. Overall, these emerging antenna technologies provide flexible and environment-adaptive solutions for next-generation wireless networks.

\section{Environment-Side Channel Reconfiguration through IRS}

As discussed in Section III, MAs provide substantial spatial DoFs for reshaping the wireless channel at the transceiver side. This capability can be further enhanced through environment-side channel reconfiguration, which introduces additional DoFs for controlling the propagation environment. In particular, in the presence of severe blockages or propagation environments with limited scattering, environment-side reconfiguration can establish additional controllable propagation paths and alleviate blockage-induced signal attenuation, complementing the channel reconfiguration achieved through transceiver-side optimization. By deploying IRSs within the propagation environment, the latter can be transformed from a passive transmission medium into a programmable network entity. This section reviews the architectures, performance advantages, and key design issues of IRS-enabled environment-side channel reconfiguration.

\begin{figure*}[!t]
	\centering
	\includegraphics[width=110mm]{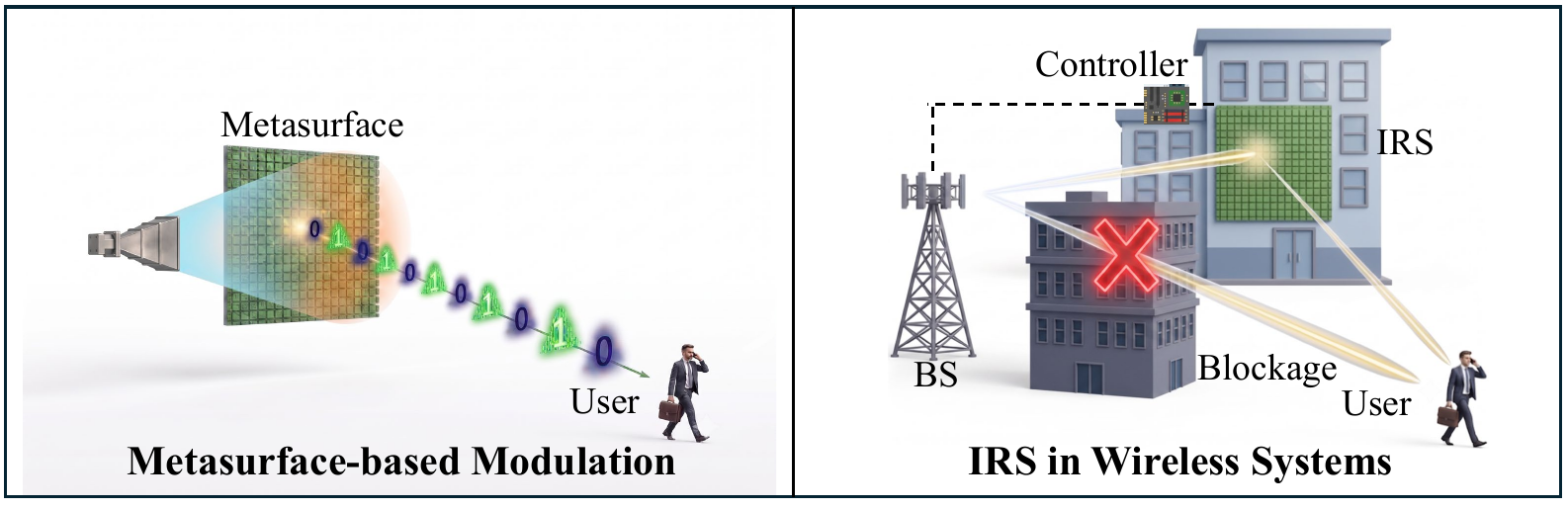}
	\caption{Architecture of metasurface and IRS in wireless systems.}
	\label{Fig_IRS_Implementations}
\end{figure*}

\subsection{Architecture}

Metasurface technology, which underpins IRS/RIS, was initially developed at the transmitter to enable spatial modulation. It was subsequently extended to the wireless propagation environment to enable environment-side channel reconfiguration. This subsection reviews the architectures of metasurface-enabled spatial modulation at the transmitter and IRS-enabled channel reconfiguration in the propagation environment.

\subsubsection{Metasurface for Spatial Modulation}

At the hardware level, the development of digital coding and programmable metasurfaces has greatly improved the electromagnetic waves control abilities \cite{Zhang2018SpaceTimeCoding,tang2020wireless}. As shown in Fig.~\ref{Fig_IRS_Implementations}, unlike conventional phased arrays that require complex and power-consuming RF chains, a programmable metasurface consists of a 2-D array of passive sub-wavelength elements integrated with simple active devices, such as PIN diodes or varactors.

By changing the bias voltage of these active devices, the reflection phase or amplitude of each unit cell can be dynamically adjusted between different discrete states. In this way, the metasurface can directly encode digital information onto the carrier wave, thereby enabling spatial modulation. As a result, programmable metasurfaces can achieve information transmission with significantly lower hardware complexity and power consumption.

\subsubsection{IRS for Channel Reconfiguration}

Based on programmable metasurfaces, IRS or RIS technology further enables wireless channel reconfiguration at the network level \cite{liaskos2018new, wu2019IRS, Huang2019RIS}. A typical IRS architecture contains a metasurface array and a smart controller, which is usually implemented using a field programmable gate array (FPGA).

As shown in Fig.~\ref{Fig_IRS_Implementations}, in this architecture, the IRS operates as a controllable reflector. The smart controller communicates with the BS or network controller to obtain CSI or optimized phase-shift configurations. According to this information, the controller adjusts the direct current (DC) bias voltages of the metasurface elements in real time. By jointly controlling the phase shifts of all reflecting elements, the IRS can strengthen the desired signals at the receiver to improve the SNR, or suppress signals in unintended directions to enhance communication security. Therefore, IRS technology can transform the wireless channel from a passive propagation environment into a programmable communication medium.

\subsection{Performance Advantages}

\begin{figure*}[!t]
	\centering
	\includegraphics[width=150mm]{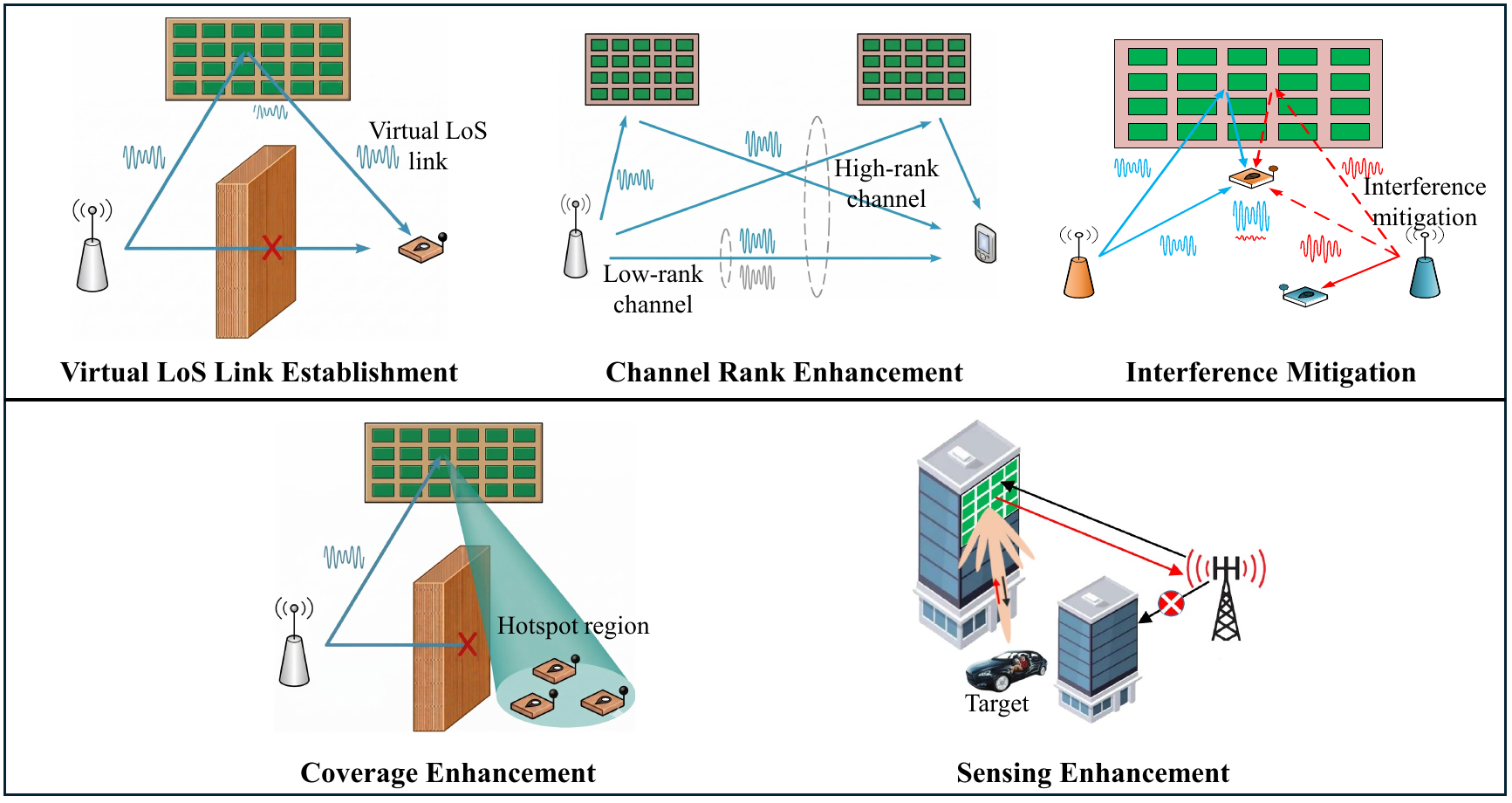}
	\caption{Performance advantages of IRS.}
	\label{Fig_IRS_advantage}
\end{figure*}

The integration of IRS into the wireless environment provides significant advantages over conventional wireless communication systems. These advantages mainly come from the unique hardware properties of IRS and their capability to actively control the wireless propagation environment for channel reconfiguration.

\subsubsection{Hardware Properties}

At the hardware level, IRSs operate with high energy efficiency \cite{liaskos2018new, wu2019IRS, Huang2019RIS}. Unlike conventional active relays, which require power-consuming RF chains and complex signal processing modules, IRS elements control electromagnetic waves using passive devices, such as PIN diodes or varactors. Therefore, IRS systems have very low power consumption. In addition, since the reflected signals are not actively amplified, IRS can avoid the additional thermal noise introduced by active relays, resulting in low noise amplification. Moreover, the reflection process is naturally simultaneous for transmission and reception, enabling full-duplex operation without the severe self-interference problem commonly encountered in conventional full-duplex transceivers.

\subsubsection{Virtual LoS Link Establishment}

In dense urban or indoor environments, obstacles often block the direct link between the transmitter and receiver, especially in millimeter wave (mmWave) and terahertz (THz) systems. By properly deploying an IRS, the reflected signals can bypass the blockage and create a virtual LoS communication path. This capability can effectively improve signal strength and maintain reliable communication in areas with severe blockage.

\subsubsection{Channel Rank Enhancement}

In environments dominated by strong LoS propagation, MIMO channels may become highly correlated, resulting in low-rank channel matrices and limited spatial multiplexing gain \cite{9849035}. IRS can intentionally introduce additional controllable reflection paths through passive beamforming. In this way, the wireless channel can be transformed into a higher-rank channel with richer channel paths, thereby improving the spatial multiplexing capability and increasing system capacity.

\subsubsection{Interference Mitigation}

The programmable phase control provided by IRS also enables efficient interference mitigation \cite{IRS25PL, wu2024ISfor6G}. In interference-limited wireless networks, the phase shifts of the reflecting elements can be optimized such that the reflected interference signals combine destructively with the direct interference signals at the receiver. This interference cancellation capability can effectively improve the SINR of the desired user without requiring complicated coordination among multiple BSs.

\subsubsection{Coverage Enhancement}

For hotspot areas with high traffic demand, IRS can provide efficient signal enhancement through passive beamforming \cite{zheng2022survey}. Instead of deploying additional small-cell BSs, the IRS can intelligently focus the reflected signal toward specific high-density user regions. This targeted signal enhancement improves the local SNR, enhances cell-edge performance, and increases the overall network throughput in a cost-effective manner.

\begin{figure}[!t]
	\centering
	\includegraphics[width=130mm]{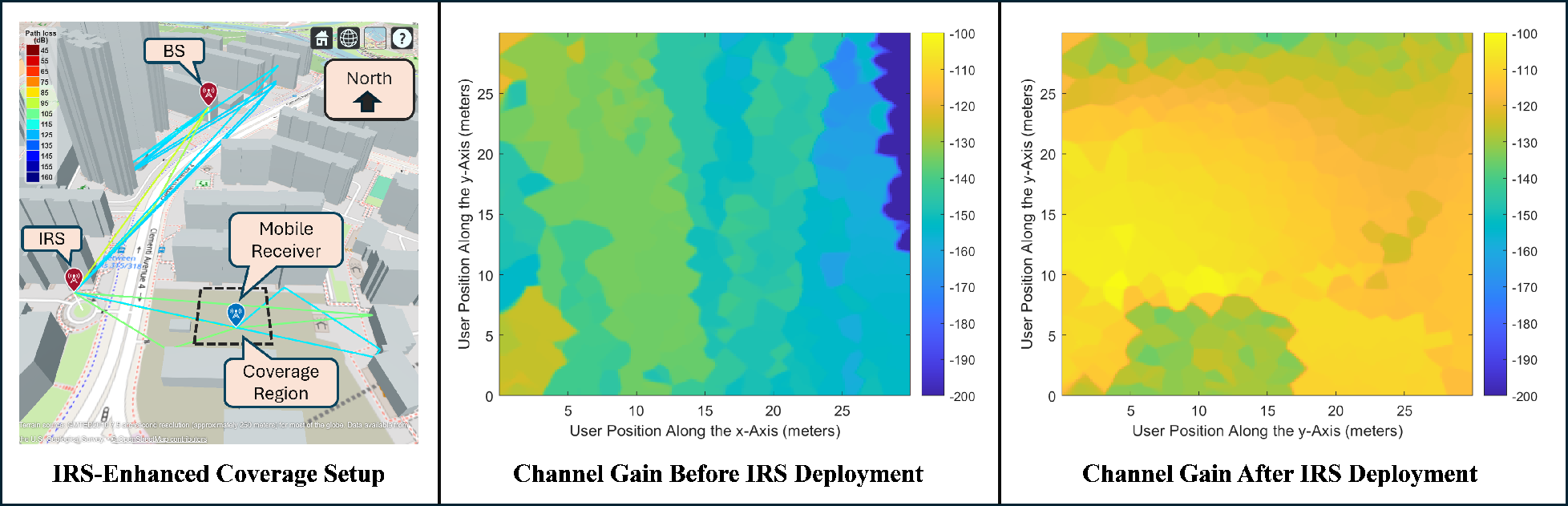}
	\caption{Wideband coverage enhancement realized via IRS deployment.}
	\label{Fig_IRS_coverage}
\end{figure}

\subsubsection{Sensing Enhancement}

Besides wireless communication, IRS can also support sensing and environmental perception \cite{Shao2022target}. By controlling the reflected electromagnetic waves, IRS can create programmable sensing paths to illuminate targets located in NLoS regions. The reflected sensing signals can then be analyzed to estimate target information such as location, velocity, and direction. Therefore, IRS is considered an important enabling technology for ISAC systems in future 6G networks.

To show the performance gain achieved by IRS deployment, Fig.~\ref{Fig_IRS_coverage} compares the average channel power gain of a wideband wireless system across the designated region before and after IRS deployment, with the detailed simulation setup described in~\cite{yan2026widebandirscoverage}. Through appropriate beam steering, the IRS directs the reflected signals toward the target coverage region, resulting in a $31$ dB improvement in the average channel power gain across the region from $-147.21$ dB to $-116.23$ dB. Meanwhile, by assuming unit transmit power, the outage probability of the receive SNR falling below $0$ dB is significantly reduced from $95.36\%$ to $11.36\%$. These results demonstrate a consistent enhancement of channel conditions throughout the region and improved the network capability to support local traffic demands.

\subsection{Design Issues}

Although IRSs provide significant theoretical advantages, their practical implementation still faces several important design challenges. Addressing these issues requires efficient deployment location selection, channel estimation, passive beamforming design, and standardization.

\subsubsection{Deployment Location Selection}

The deployment location selection is a critical design issue for IRS-assisted wireless systems, as it directly determines the availability and quality of the reflected propagation paths. In practical deployments, IRSs are typically installed at a set of candidate sites with favorable propagation conditions and sufficient physical space. The key problem is therefore to select an appropriate subset of candidate sites to improve network coverage while balancing deployment cost and communication performance. In~\cite{FU2024MultiIRS}, a multi-IRS deployment framework was developed based on large-scale channel knowledge, where the candidate sites are jointly optimized with the heights, orientations, and numbers of reflecting elements of the deployed IRSs. By characterizing the average cascaded channel power gain between the BS and different coverage grids, the IRS deployment problem was formulated to minimize the total deployment cost subject to a prescribed coverage requirement~\cite{FU2024MultiIRS}. For more complex environments with dense blockages, distributed multi-IRS deployment can further establish cascaded LoS paths and improve coverage. In~\cite{FU2025MultiIRS}, the considered region is divided into multiple cells with predetermined candidate IRS locations, and an optimal subset of candidate locations is selected jointly with the number of reflecting elements at each selected site. In particular, both passive and active IRSs are considered, and the deployment design aims to satisfy a target SNR over all cells while minimizing the overall deployment cost~\cite{FU2025MultiIRS}. These studies show that, under fixed infrastructure constraints, IRS location selection can effectively reshape the propagation topology and improve coverage by strategically placing IRSs at propagation-favorable sites.

The above approaches mainly consider infrastructure-level deployment with fixed candidate sites. In dynamic wireless environments, where blockages, users, or propagation conditions vary significantly over time, mobile IRSs mounted on UAVs or ground vehicles provide an additional DoF by dynamically adjusting their locations~\cite{10028753}. In such systems, the IRS position can be jointly optimized with its reflection coefficients according to the instantaneous or large-scale channel conditions, offering greater flexibility than conventional fixed deployment.

\subsubsection{Cascaded Channel Acquisition}

Accurate CSI is essential for designing effective IRS reflection coefficients. However, since passive IRSs do not contain active RF chains, they cannot directly estimate wireless channels. One possible solution is to adopt semi-passive IRS architectures, where a small number of active sensing elements are integrated into the surface to estimate channel parameters such as AoAs and path gains~\cite{9340586}. Although this approach improves channel estimation accuracy, it increases hardware complexity and power consumption.

An alternative approach is to estimate the cascaded transmitter-IRS-receiver channel at the active transceivers through pilot training~\cite{zheng2020IRSOFDM}. Early methods employed ON/OFF reflection patterns to separately estimate the contribution of each IRS element. To reduce training overhead and improve estimation efficiency, later studies proposed full-ON training schemes and element-grouping methods~\cite{zheng2020IRSOFDM}. Furthermore, since the cascaded channel dimension can be very large, compressed sensing techniques~\cite{9354904} and hierarchical codebook-based feedback methods~\cite{9367208} have been developed to reduce pilot overhead.

Despite these advances, conventional pilot-based channel estimation still incurs substantial training overhead and often requires modifications to existing communication protocols. To address this limitation, recent studies have proposed a new channel acquisition paradigm based solely on received signal power measurements, such as the reference signal received power (RSRP) \cite{yan2025power, sun2024power, 10907801, yan2026widebandirscoverage}. Since RSRP measurements are readily available in existing cellular and wireless local area network (WLAN) systems, these methods exploit user-side power measurements collected under different IRS reflection configurations to estimate the channel autocorrelation matrix or directly infer the channel using neural networks. By eliminating the need for explicit instantaneous CSI estimation and dedicated pilot transmission, this framework substantially reduces training overhead while remaining fully compatible with existing wireless communication protocols.

\subsubsection{Reflection Coefficient Optimization}
After obtaining CSI, the transmit beamforming at the BS and the reflection coefficients of the IRS must be jointly optimized to maximize communication performance~\cite{8970580}. However, perfect CSI is difficult to obtain in practice because of channel estimation errors, feedback delays, and channel variations. Therefore, robust beamforming methods that operate under imperfect or statistical CSI have been extensively studied~\cite{9110587}.

Building upon recent advances in power-measurement-based channel acquisition, IRS reflection coefficient optimization can be performed directly using the estimated channel autocorrelation matrices~\cite{yan2025power, sun2024power, 10907801, yan2026widebandirscoverage}. Rather than relying on full instantaneous CSI, the reflection coefficients are optimized to maximize the average channel power gain. Such statistical optimization reduces the reliance on instantaneous CSI acquisition and is particularly attractive for practical IRS deployments.

Moreover, CSI-free blind beamforming has emerged as a promising approach for practical IRS control~\cite{ren2022configuring}. Instead of explicitly estimating the reflected channels, the IRS reflection coefficients are optimized directly based on received signal power measurements under different reflection configurations. This approach is particularly appealing for existing wireless networks, as it does not require CSI acquisition or coordination with the BS, and can operate in a plug-and-play manner. Its practical feasibility has been demonstrated through field tests in a real-world 5G network, where a 256-element IRS with four discrete phase-shift states was deployed for 2.6-GHz downlink transmission. The IRS operated without BS-side knowledge or coordination and achieved substantial improvements in received signal strength and SINR, demonstrating the potential of blind beamforming for practical large-scale IRS deployment.

\subsubsection{Standardization}

As environment-side channel reconfiguration advances from theoretical research toward practical deployment, standardization has become the key enabler. The 3GPP has actively investigated network-controlled electromagnetic nodes. In particular, 3GPP Release 18 introduced the concept of network-controlled repeaters (NCRs) to enhance network coverage and capacity in a coordinated manner \cite{guo2022comparison}. Unlike conventional repeaters, which indiscriminately amplify both desired signals and interference, NCRs receive control information (e.g., beamforming configurations and ON/OFF switching patterns) directly from the BS to enable network-coordinated operation. Although NCRs employ active RF components that are different from fully passive IRSs, they share similar deployment strategies, beam management mechanisms, and network-controlled operating principles with IRSs. Consequently, the standardization of NCRs establishes an important regulatory and architectural foundation for the future integration of IRSs into standardized 6G networks.

\subsection{Other Relevant Technologies}

The intelligent surface concept has evolved into several advanced architectures that provide additional flexibility and performance gains for future 6G networks. To fully exploit spatial DoFs, dynamic metasurface antennas (DMAs) \cite{pulido2016application} and reconfigurable holographic surface (RHSs) \cite{deng2021RHS} combine tunable metamaterial elements with feeding waveguides to perform active transmission. Compared with conventional phased arrays, these architectures can achieve high aperture efficiency with reduced hardware complexity. Other recent developments include stacked intelligent metasurfaces, which employ multiple programmable layers for wave-domain signal processing \cite{an2024stacked}.

At the same time, several new IRS/RIS architectures have been developed to overcome the limitations of conventional reflecting surfaces. Traditional IRSs can only control signals within one side of the surface. To provide full-space coverage, simultaneously transmitting and reflecting surfaces \cite{mu2021simultaneously} and intelligent omni-surfaces \cite{zhang2021intelligentO} have been proposed, allowing both transmission and reflection of incident signals. Another important development is the beyond-diagonal IRS/RIS \cite{li2022beyond}, which exploits coupling among reflecting elements and removes the conventional diagonal reflection matrix constraint, thereby providing greater design flexibility and improved performance. More recently, flexible intelligent metasurfaces have been proposed \cite{an2025flexible}, which flexibly adjust the position of each element to adapt to changing channel conditions. Moreover, by optimizing not only the reflection coefficients but also the physical position and orientation of the surface, movable intelligent surfaces introduce additional spatial DoFs and provides new opportunities for proactive wireless environment reconfiguration \cite{zheng2025movableRIS}.

\section{Future Perspectives}
Despite the substantial advances in channel cognition and reconfiguration, the paradigm of channel cognition and reconfiguration remains at an early stage of development. As 6G wireless networks continue to evolve, there is an increasing need to move beyond isolated technological advances toward unified, scalable, and adaptive frameworks that seamlessly integrate channel cognition and reconfiguration. To fully harness the potential of this paradigm and inspire future innovations, this section discusses several promising research directions.

\begin{figure}[!t] 
	\centering 
	\captionsetup[subfloat]{captionskip=5pt} 
	\subfloat[Illustration of full-spectrum channel cognition in 3-D space-air-ground-sea integrated networks.]{
		\includegraphics[width=0.8\textwidth]{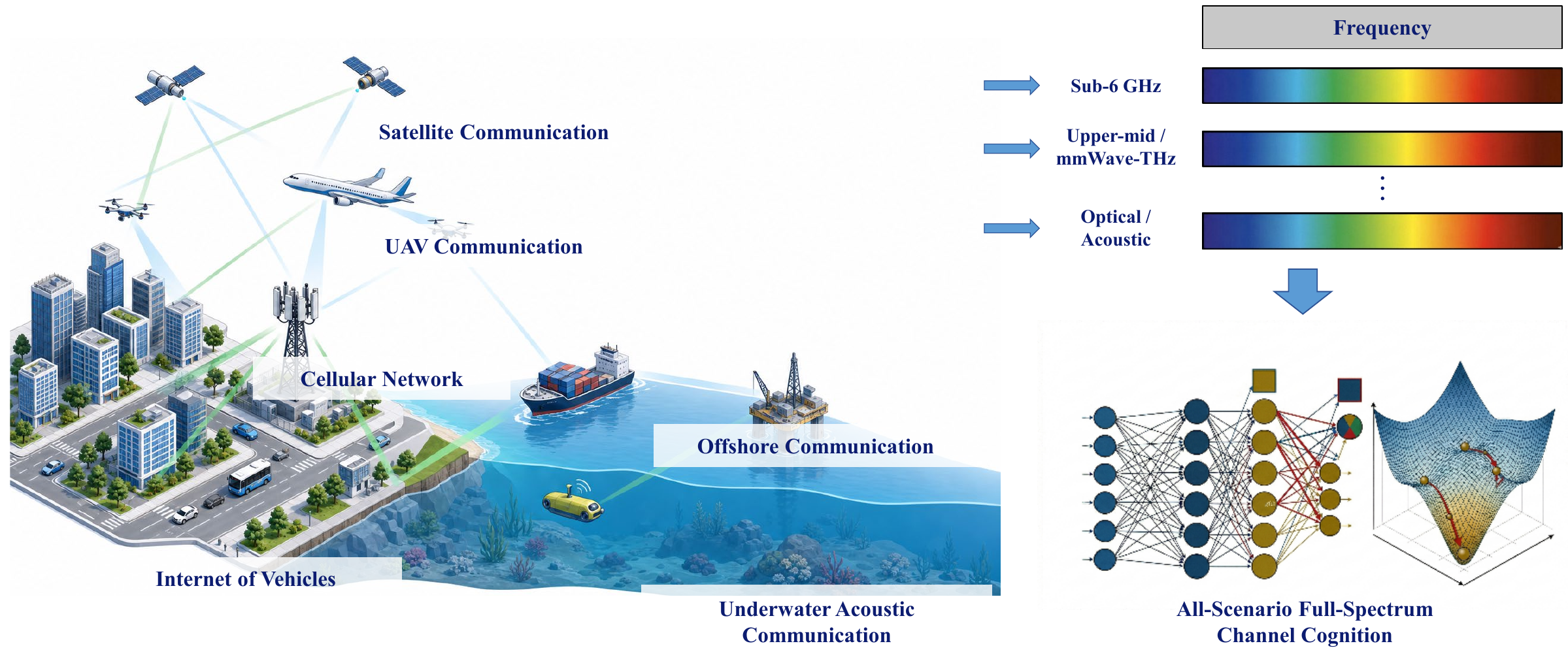}
		\label{Fig_all_scenario}
	}\hspace{0.2cm} 
	\subfloat[Collaborative channel reconfiguration via diverse technologies at both transceiver and environment sides.]{
		\includegraphics[width=0.8\textwidth]{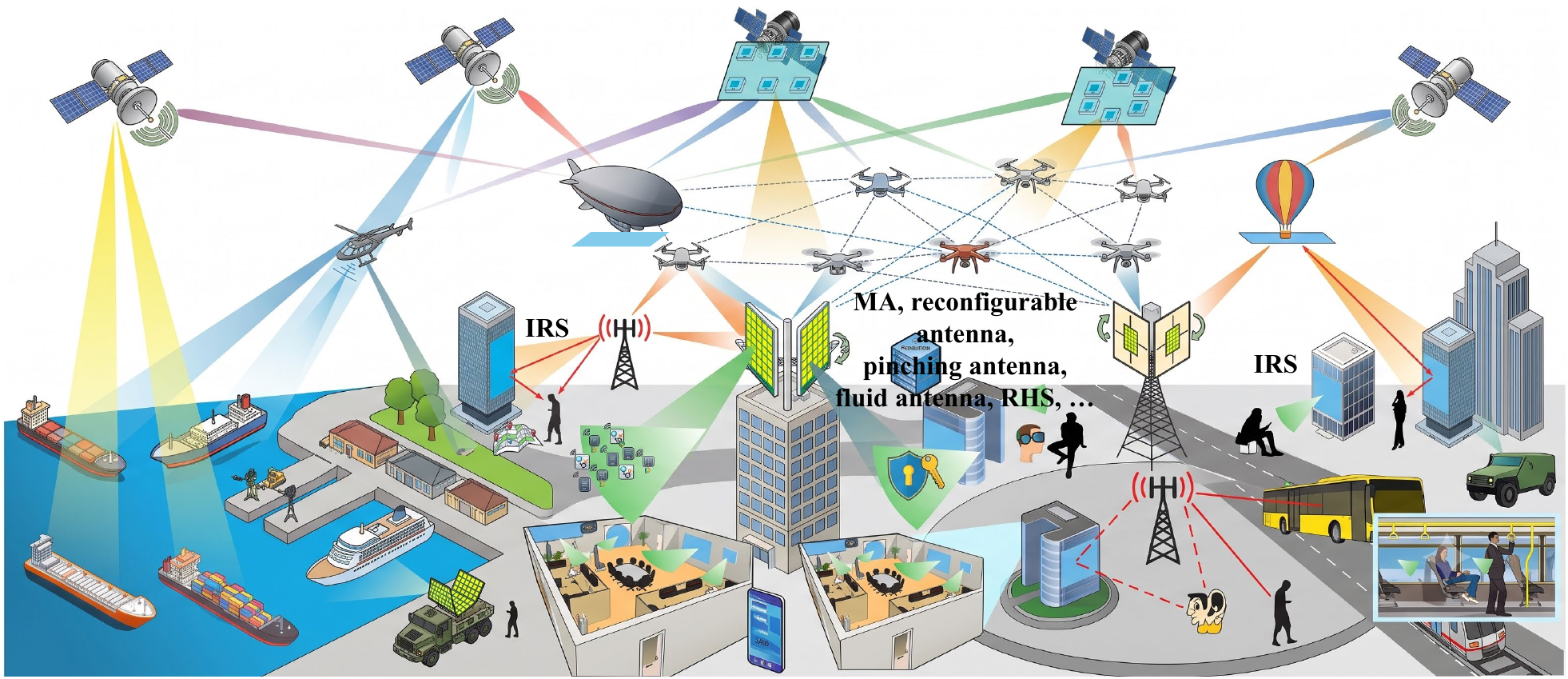}
		\label{Fig_Future_B}
	}\hspace{0.2cm} 
	\subfloat[Embodied AI network architecture for closed-loop channel cognition and reconfiguration.]{
		\includegraphics[width=0.8\textwidth]{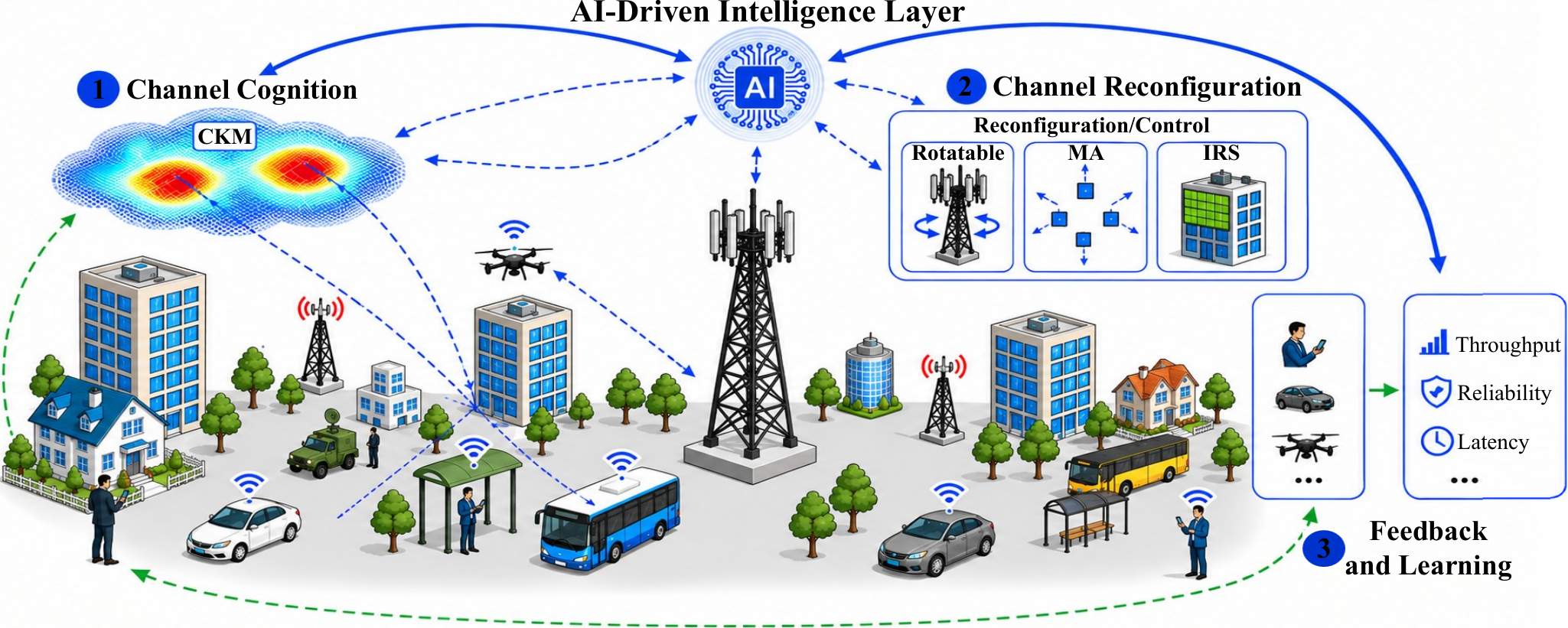} 
		\label{Fig2}
	}
	\caption{Representative future perspectives for channel cognition and reconfiguration in the 6G era.}
	\label{Fig_Future}
	\vspace{-0.26cm} 
\end{figure} 

\subsection{Full-Spectrum Channel Cognition in 3-D Space-Air-Ground-Sea Integrated Networks}

As illustrated in Fig.~\ref{Fig_all_scenario}, 6G networks are expected to integrate terrestrial, aerial, maritime, underwater, and non-terrestrial segments into a unified 3-D space-air-ground-sea architecture. Representative use cases include terrestrial cellular and Internet-of-vehicles communications, offshore and ship-to-shore links, underwater acoustic communications, low-altitude and high-altitude communications, and satellite networks. These segments exhibit markedly different propagation conditions, mobility patterns, operating altitudes, and frequency bands. Channel cognition must therefore extend beyond a single scenario or frequency band to provide unified awareness across heterogeneous domains.

Such integration introduces two closely related requirements. First, full-spectrum cognition must cover sub-6 GHz, upper-mid-band, mmWave, THz, optical, and underwater acoustic bands. Because propagation mechanisms, antenna configurations, and channel statistics vary substantially across these bands, a CKM designed for a single band cannot support integrated network management. Second, channel knowledge must be represented in 3-D and across domains. Whereas terrestrial links can generally be characterized by 2-D or 3-D CKMs, UAVs and high-altitude platforms (HAPs) require 3-D airspace representations; maritime users introduce sea-surface and offshore propagation effects; and satellite links extend the cognition space to space-air and space-ground channels. Future CKMs should therefore describe location-specific channel knowledge over wider spatial and spectral domains while retaining the characteristics of individual network segments.

A key research direction is the development of unified CKM frameworks for cross-scenario, full-spectrum channel cognition. Cross-domain CKMs should fuse terrestrial, aerial, maritime, underwater, and satellite measurements, whereas spectrum-aware CKMs should capture band-dependent propagation features. Efficient update mechanisms are also needed for highly dynamic nodes, such as vehicles, vessels, UAVs, HAPs, and low-Earth-orbit (LEO) satellites. Cross-band CKM construction can exploit frequency-domain correlations for knowledge transfer and spatial sparsity for high-dimensional map recovery from limited samples, while explicitly accounting for band-specific scattering, blockage, and material responses that restrict direct transfer. The integration of sensing, communication, and environmental information can further facilitate CKM construction and maintenance in sparsely observed or difficult-to-measure regions, such as offshore and underwater environments. Finally, standardized interfaces and query mechanisms are required to support the joint use of domain-specific CKMs for coverage planning, mobility management, beam coordination, and proactive resource allocation in integrated 6G networks.

\subsection{Collaborative Channel Reconfiguration via Diverse Technologies}

As illustrated in Fig.~\ref{Fig_Future_B}, the traditional paradigm of passive channel adaptation will gradually shift toward active collaborative channel reconfiguration via diverse technologies. Future 6G systems will no longer simply adapt to a random and uncontrollable propagation environment. Instead, they will actively shape the wireless medium to improve communication and sensing performance \cite{ma2026survey}. This transformation will rely on the integration and cooperation of several emerging channel reconfiguration technologies. Based on the channel information from channel cognition, heterogeneous spatial and electromagnetic DoFs can be jointly utilized and optimized. For example, the wireless propagation environment can be dynamically controlled through IRSs/RISs, while the transceiver side can achieve greater flexibility through MAs, fluid antennas, and reconfigurable antenna technologies.

A key research direction is the development of unified and low-overhead optimization frameworks that can efficiently coordinate these technologies. This includes jointly utilizing channel cognition, integrated sensing information, and wireless environment knowledge to optimize the configurations of transceiver- and environment-side components. Through such collaborative design, future networks can further improve channel capacity, coverage, and sensing accuracy while maintaining acceptable hardware complexity and implementation cost.

\subsection{Embodied AI Networks for Closed-Loop Channel Cognition and Reconfiguration} 

A promising direction is to develop intelligent networks for channel cognition and reconfiguration, where channel cognition, intelligent reconfiguration, feedback evaluation, and self-learning operate collaboratively within a unified closed loop. As illustrated in Fig. \ref{Fig2}, the network first acquires environment-aware channel knowledge from multi-source observations, including network states, user distributions, sensing information, and propagation measurements. Such knowledge is then exploited to coordinate reconfigurable resources at the transceiver and propagation-environment sides, thereby improving the effective channel conditions. The post-reconfiguration measurements and performance feedback are further used to update channel knowledge, refine prediction models, and optimize subsequent decisions. Through this closed-loop mechanism, wireless networks are expected to evolve from passively adapting to channels toward embodied AI systems capable of proactively cognizing, reconfiguring, and continuously optimizing wireless channels. Future research is still needed on low-overhead knowledge updating, reliable feedback modeling, cross-timescale control, safe learning, and collaborative optimization of heterogeneous reconfigurable resources.

\section{Conclusions}  \label{Sec_conclusion}

In this paper, we have provided a comprehensive overview of the emerging paradigm shift from conventional passive channel adaptation toward proactive channel cognition and reconfiguration in the 6G era. We first introduced channel cognition through CKMs, which offer a systematic framework for learning, representing, predicting, and utilizing channel knowledge across spatial, temporal, and frequency domains. The fundamental definitions, construction methods, and representative applications of CKMs were comprehensively reviewed. Building upon channel cognition, we further reviewed channel reconfiguration technologies from both transceiver and propagation-environment perspectives. Specifically, MAs enable transceiver-side channel reconfiguration through adaptive antenna positioning/rotation, while IRSs facilitate environment-side channel reconfiguration by intelligently reconfiguring electromagnetic wave propagation. Their system architectures, performance advantages, and key design issues were systematically discussed. Despite the significant progress achieved in recent years, channel cognition and reconfiguration remain in their early stages of development. We hope this paper will serve as a valuable resource for researchers and practitioners, inspiring further innovations to unlock the full potential of these promising technologies in realizing intelligent and adaptive wireless networks.


\bibliographystyle{scis}
\bibliography{IEEEexample}



\end{document}